\documentclass[reprint,amsmath,amssymb,aps,nofootinbib,superscriptaddress,prx,floatfix]{revtex4-2}
\IfFormatAtLeastTF{2023-06-01}{\let\label\kernelLabel}{}
\usepackage{amsmath,amssymb,amsthm}
\usepackage[cjk]{kotex}
\usepackage{graphicx}
\usepackage[dvipsnames]{xcolor}
\usepackage[normalem]{ulem}
\usepackage{microtype}
\usepackage[colorlinks, linkcolor=Red, citecolor=NavyBlue, urlcolor=NavyBlue]{hyperref}
\usepackage[all]{hypcap} 
\usepackage{tikz}
\usetikzlibrary{positioning}
\usepackage{enumitem}
\usepackage{physics}
\usepackage{bm}
\usepackage{dsfont}
\usepackage{mathtools}
\renewcommand{\v}{\textbf}

\begin{document}

\title{Long-Range Toric Code: Finite-Temperature Topological Order in Two Dimensions}

\author{Jacopo Gliozzi}\affiliation{Department of Physics and Center for Theory of Emergent Quantum Matter, Pennsylvania State University, University Park, Pennsylvania 16802, USA}
\author{Taylor L. Hughes} 
\affiliation{Department of Physics and Anthony J. Leggett Institute for Condensed Matter Theory, University of Illinois  Urbana-Champaign, 1110 West Green Street, Urbana, Illinois 61801, USA} 
\author{Kyung-Su Kim (김경수)}\email{kyungsu@illinois.edu}
\affiliation{Department of Physics and Anthony J. Leggett Institute for Condensed Matter Theory, University of Illinois  Urbana-Champaign, 1110 West Green Street, Urbana, Illinois 61801, USA} 

\date{\today}

\begin{abstract}
We introduce two-dimensional toric code models with additional long-range loop interactions. Our models preserve the topologically ordered ground states of the ordinary toric code, while confining anyons at zero temperature. 
For sufficiently long-ranged interactions, we argue that this confinement persists at finite temperatures, strongly constraining anyon mobility and
giving rise to finite-temperature topological order.
This topological phase acts as a self-correcting quantum memory and exhibits long-range entanglement. 
For shorter-ranged interactions, signatures of topological order survive only up to an exponentially large length scale at finite temperature, yielding a partially self-correcting quantum memory.
\end{abstract}
\maketitle

\section{Introduction}\label{sec:introduction}

Topologically ordered phases of matter in two dimensions (2D) can act as a quantum memory at zero temperature, with quantum information encoded nonlocally in a degenerate ground state manifold~\cite{Kitaev2003Jan}. The topological order of the toric code is a 
classic example
and has recently been prepared on programmable quantum devices~\cite{Satzinger2021, Semeghini2021}. 
At finite temperatures, however, thermally excited anyons rapidly destroy the stored information: 
separating two anyons across a system of size $L$ costs a finite energy $\Delta E$, but leads to an entropy gain $\Delta S \sim \ln L$. The free energy, $\Delta F = \Delta E - T \Delta S$, therefore favors anyon proliferation at any temperature $T > 0$, destabilizing the 
2D topological order
and its associated quantum memory~\cite{Hastings2011,Nussinov2008Feb,Bravyi2009nogo,Alicki2009,Yoshida2011, Poulin2013, Weinstein2019}.

There is an analogous situation for phases characterized by spontaneously broken global symmetries, which can store {\it classical} information. For example, a 2D Ising ferromagnet acts as a classical memory, with the sign of the magnetization encoding one bit. In one dimension, however, the Ising model does not show spontaneous ferromagnetism at any nonzero temperature. 
This is because domain wall excitations are deconfined, {much like anyons in 2D}: domain walls cost a finite energy, but have a logarithmically growing entropy $\Delta S \sim \ln L$.
The proliferation of domain walls for $T>0$ then results in a vanishing magnetization {and destabilizes the classical memory}. 

The finite-temperature instability of 1D ferromagnetism can be remedied by adding long-range Ising interactions whose strength decays with the separation $r$ between two spins as $\sim 1/r^\alpha$. 
In this case, a pair of domain walls separated by a distance $d$ has energy
\begin{align}
\label{eq:domain_wall_energetics}
    \Delta E_{\rm dw} (d) 
    &\sim 
    \begin{cases}
    d^{2-\alpha} & (1 < \alpha < 2) 
    \\
    \ln d & (\alpha = 2) 
    \\
    \text{const.}
    & (\alpha > 2)
    \end{cases}
    .
\end{align}
Consequently, when $1 < \alpha \leq 2$, the long-range interactions confine domain walls and stabilize a ferromagnetically ordered phase below a critical temperature~\cite{Dyson1969,thouless1969long, frohlich1982phase}.

Inspired by this analogy, we construct a class of long-range toric code (LRTC) models, whose terms consist of large-loop stabilizers with coefficients that decay with loop perimeter, in addition to the usual star and plaquette terms. The resulting eigenstates, and in particular the topological ground state manifold, remain identical to those of the ordinary toric code. However, the energies of excited states are dramatically reorganized. For example, for some ranges of parameters the energy cost of separating two anyons grows without bound in the LRTC, thus confining the anyon excitations.

This leads to an unusual situation in which a topologically ordered ground state coexists with confined anyons, in contrast to local Hamiltonians, where ground state topological order and anyon deconfinement go hand in hand. Thus, our long-range toric code models exhibit zero-temperature ``confining topological order.'' Confinement also suppresses the thermal proliferation of anyons, allowing topological order to persist to nonzero temperatures. As in the Ising model, our long-range approach is complementary to the higher-dimensional route to finite-temperature topological order: 
in the 4D toric code, for example, the relevant excitations are loops instead of pointlike anyons, and have an energy cost that grows with their size~\cite{Dennis2002, Alicki2012}.
Both approaches energetically penalize spatially extended defect configurations, thereby suppressing their entropic proliferation.

Our article is organized as follows. After introducing our long-range toric code models in Sec.~\ref{sec:LRTC}, we discuss the effective confining interactions between isolated anyons in Sec.~\ref{sec:anyon_energetics}.
We then examine how this confinement is modified by a thermal background in Sec.~\ref{sec:energetics_finiteT}. In one regime, confinement persists at low temperatures, stabilizing finite-temperature topological order, while in another regime it is screened beyond an exponentially large length scale. 
Monte Carlo simulations in Sec.~\ref{sec:transition} provide evidence for a thermodynamic transition in the former regime, and a sharp finite-size crossover in the latter, consistent with the analytical arguments of Sec.~\ref{sec:energetics_finiteT}.

We then characterize the information-theoretic properties of these regimes. In Sec.~\ref{sec:memory}, we provide analytical and numerical evidence that the fully confining regime has finite-temperature topological order and 
acts as a passive quantum memory, storing quantum information for a time that diverges as $L\rightarrow \infty$. In contrast, the partially confining regime gives rise to a partially self-correcting quantum memory, whose protection increases with system size only up to a finite length scale. In Sec.~\ref{sec:lre}, we argue that low-temperature Gibbs states in the fully confining regime are long-range entangled and discuss finite-size signatures of long-range entanglement in the partially confining regime. 
Finally, in Sec.~\ref{sec:discussion}, we summarize the work and discuss directions for future research.

\begin{figure}[t!]
\centering
\includegraphics[width=\linewidth]{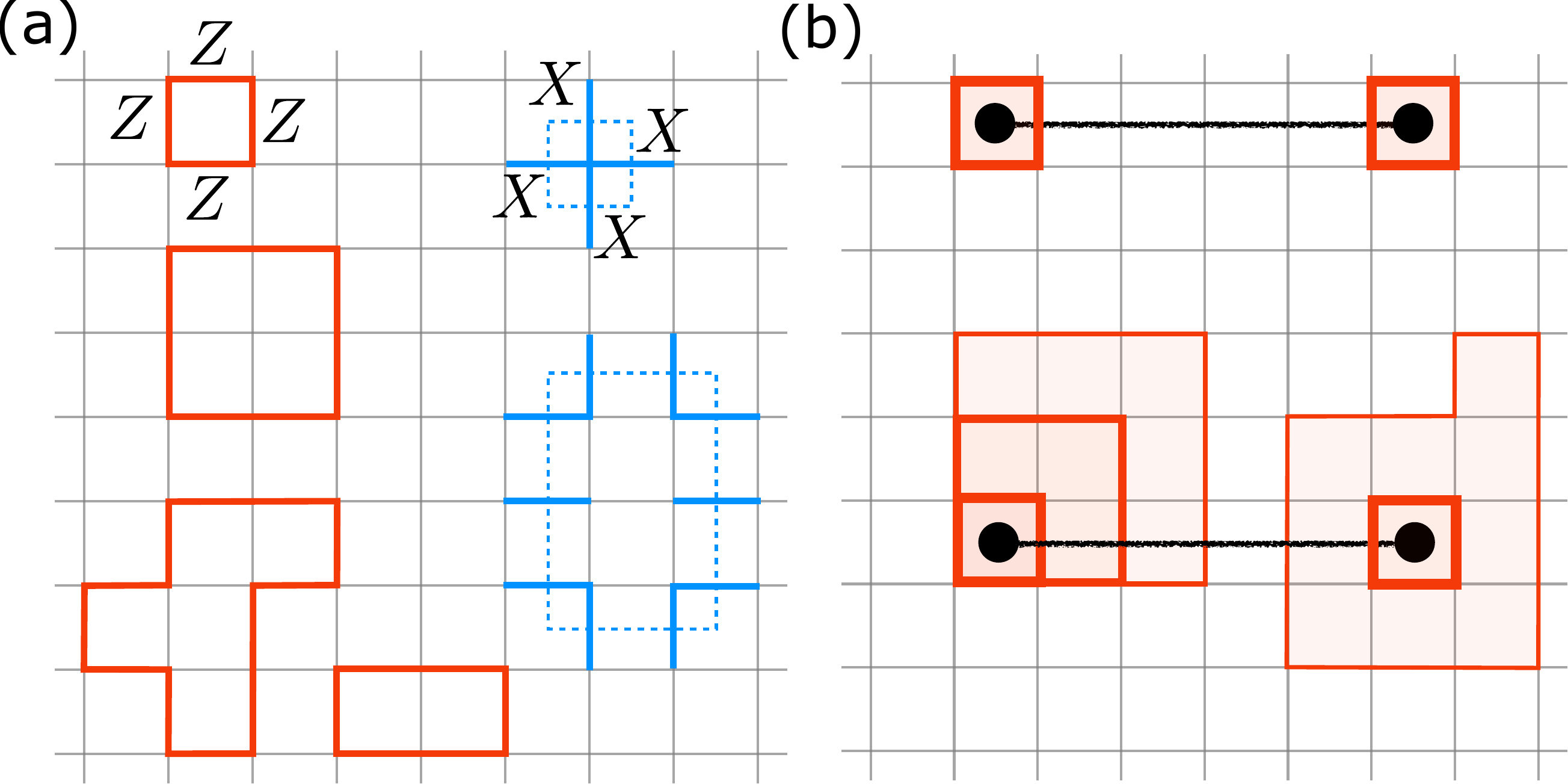}
\caption{(a) Top: Plaquette (red) and star (blue) terms of the toric code form squares on the direct and dual lattices, respectively. Bottom: 
The long-range toric code includes 
additional stabilizers supported on larger loops. 
In the main text, we focus on the square-loop LRTC model. More general loops are discussed in Appendix \ref{app:other_loops}.
(b) A pair of magnetic anyons connected by a flux string. The energy of this configuration is determined by the loops that enclose exactly one anyon. This energy is bounded in the ordinary toric code (top), but increases with anyon separation in the LRTC (bottom).
} 
\label{fig:allowed}
\end{figure}

\section{Long-range toric code models}
\label{sec:LRTC}
We now introduce a family of modified toric code Hamiltonians with long-range loop interactions. In the ordinary toric code, Hamiltonian terms are products of Pauli operators $X, Z$ around elementary square loops on the lattice ($B_p$) and the dual lattice ($A_v$):
\begin{equation}\label{eq:ham_tc}
\begin{gathered}
    H_\text{tc} = - J_B \sum_{p} B_p -  J_A \sum_v A_v,\\
     B_p = \begin{tikzpicture}[scale = 0.5, baseline = {([yshift=-.5ex]current bounding box.center)}]
        \draw[color = black] (-1, -1) -- (-1, 1) -- (1, 1) -- (1, -1) -- cycle;
        \node at (-1, 0) {\small $Z$};
        \node at (1, 0) {\small $Z$};
        \node at (0, -1) {\small $Z$};
        \node at (0, 1) {\small $Z$};
        \node at (0, 0) {\small $\textcolor{black}{p}$};
    \end{tikzpicture},
    \quad 
    A_v = \begin{tikzpicture}[scale = 0.5, baseline = {([yshift=-.5ex]current bounding box.center)}]
    \draw[black] (1.5,0) -- (-1.5, 0);
    \draw[black] (0,1.5) -- (0,-1.5);
    \node at (0.9, 0) {\normalsize $X$};
    \node at (-0.9, 0) {\normalsize $X$};
    \node at (0, 0.9) {\normalsize $X$};
    \node at (0, -0.9) {\normalsize $X$};
    \node at (0.2, 0.2) {\small $\textcolor{black}{v}$};
    \end{tikzpicture}.
\end{gathered}
\end{equation}
The plaquette and star terms in Eq.~\eqref{eq:ham_tc} are mutually commuting stabilizers, and the  ground states satisfy $A_v = B_p = 1$ on every vertex $v$ and 
plaquette $p$. 
In our long-range models, we supplement these elementary stabilizers with larger-loop stabilizers whose coefficients decay with loop perimeter.
We primarily focus on square-loop interactions; other choices of loop ensembles, including rectangles and self-avoiding polygons, are discussed in Appendix~\ref{app:other_loops}. 
Some examples of larger loop interactions are depicted in Fig.~\ref{fig:allowed}(a).

Concretely, the interactions in the square-loop LRTC model are a set of $Z$-type ($X$-type) square loops  on the direct (dual) lattice, denoted by $\mathcal{S}$ ($\tilde{\mathcal{S}}$). For each square $\mathfrak s \in \mathcal S$, 
the corresponding $Z$-type stabilizer term is
\begin{align}
    \mathcal B_{\mathfrak s} 
    = \prod_{i\in \partial \mathfrak s} Z_i
    = \prod_{p\in {\rm int}(\mathfrak s) } B_{p}
    ,
\end{align}
where $\partial \mathfrak s$ is the set of links along the boundary of $\mathfrak s$ and  ${\rm int} (\mathfrak s)$ is the set of elementary plaquettes $p$ that are contained inside $\mathfrak s$. The $X$ stabilizer terms $\mathcal A_{\tilde {\mathfrak s}}$ are defined analogously for square loops on the dual lattice, $\tilde{ \mathfrak {s}} \in \tilde {\mathcal S}$.
The operators $\mathcal B_{{\mathfrak s}}$ and $\mathcal A_{\tilde {\mathfrak s}}$ are thus contractible Wilson and  't Hooft loop operators, respectively. 
Assigning coupling strengths that decay algebraically with loop perimeter, the Hamiltonian becomes
\begin{equation}
\label{eq:ham_square}
    H_\text{sq} = 
    -\sum_{\mathfrak s \in \mathcal{S}} 
    \frac{J_B}{
    \ell(\mathfrak s)
    ^{\alpha_B}} \mathcal B_{\mathfrak s} 
    - \sum_{\tilde{\mathfrak s} \in \tilde{\mathcal{S}}}
    \frac{J_A}{\ell(\tilde{\mathfrak s})^{\alpha_A}} \mathcal A_{\tilde{\mathfrak s}},
\end{equation}
where $\ell(\mathfrak s)=4R$ is the perimeter of 
$\mathfrak s$, $R$ is the side length of $\mathfrak s$, $J_{A, B}>0$ are coupling constants, and $\alpha_{A, B}$ determine the range of interactions.
For simplicity, we will mostly work with $J_A = J_B = 1$ and $\alpha_A = \alpha_B = \alpha$, so that the model has an exact electric-magnetic duality exchanging the direct and dual lattices.

Since the large-loop stabilizers are products of elementary star or plaquette operators, $H_{\rm sq}$ shares a common eigenbasis with the ordinary toric code, labeled by the eigenvalues of the elementary stabilizers $A_v,B_p=\pm1$ and the global topological sector. Because all the couplings are positive, the two models have identical ground states and hence the same ground state topological order. However, the excited-state energies are {drastically} reorganized by the additional long-range interactions. As we show in the next section, for a certain range of $\alpha$, {the energy cost of separating two anyons by a distance $d$ diverges as $d\to\infty$.}
In contrast to conventional local Hamiltonians, where ground state topological order and anyon deconfinement go hand in hand, we refer to this unusual regime as zero-temperature ``confining topological order.'' 
The confinement of anyons then allows the topological order and its associated quantum memory to persist to finite temperatures.

Nonlocal interactions are also used to stabilize finite-temperature topological order in quantum low-density parity-check (qLDPC) codes on expander graphs. Such systems have Hamiltonians that are $k$-local but cannot be naturally embedded in a Euclidean lattice~\cite{Breuckmann2021,Rakovszky2023,Hong2025ldpc,Placke2024ldpc}. In contrast, our LRTC model is defined on a 2D Euclidean lattice, but relaxes $k$-locality by introducing large-loop stabilizers. 
Despite its long-range interactions, the energy bandwidth of $H_\text{sq}$ 
is extensive for $\alpha >1$: 
\begin{equation}\label{eq:bandwidth}
    E_{\rm tot}\sim L^2\sum_{R=1}^L\frac{1}{(4R)^\alpha} \sim L^2,
\end{equation}
where the prefactor counts the $L^2$ possible positions of a square of fixed size, and the sum is over all possible sizes $R$. Thus, the LRTC is a thermodynamically well-defined 2D framework for exploring finite-temperature topological order.

We remark that our model differs from earlier 
long-range toric code models with pairwise interactions between 
local anyon occupations or stabilizer variables in two respects~\cite{Hamma2009,chesi2010self,hutter2012self,Brown2016Nov, wootton2013topological,Landon-Cardinal2015}. First, the proposed quantum memory in the earlier models relies on interactions that are nonsummable in two dimensions, resulting in a super-extensive energy spectrum. This contrasts with the extensive LRTC Hamiltonian discussed above. Second, our construction instead makes the gauge-field interactions themselves long-ranged by extending the elementary plaquette and star terms to Wilson and 't Hooft loops of arbitrary size. This parallels the long-range Ising model, where the fundamental short-range spin-spin interaction is extended to act over arbitrary distances. In this sense, the LRTC provides a natural gauge-theoretic way to introduce long-range interactions into the toric code.

\section{Anyon energetics at $T=0$}\label{sec:anyon_energetics}

To preserve topological order at finite temperature, anyons must remain confined. From the aforementioned free energy argument, confinement occurs if the energy cost of separating anyons outweighs the entropy gained by their proliferation. Starting from a ground state, the fundamental instability amounts to nucleating two anyons and separating them by an $\mathcal{O}(L)$ distance. The separated anyons have an entropy of $\Delta S \sim \ln L$, so the energy cost of their separation must scale at least like $\mathcal{O}(\ln L)$ to confine them. In the long-range toric code, calculating the energy cost of an anyon configuration reduces to a loop enumeration problem that we describe below.

\begin{figure}[t!]
\centering
\includegraphics[width=\linewidth]{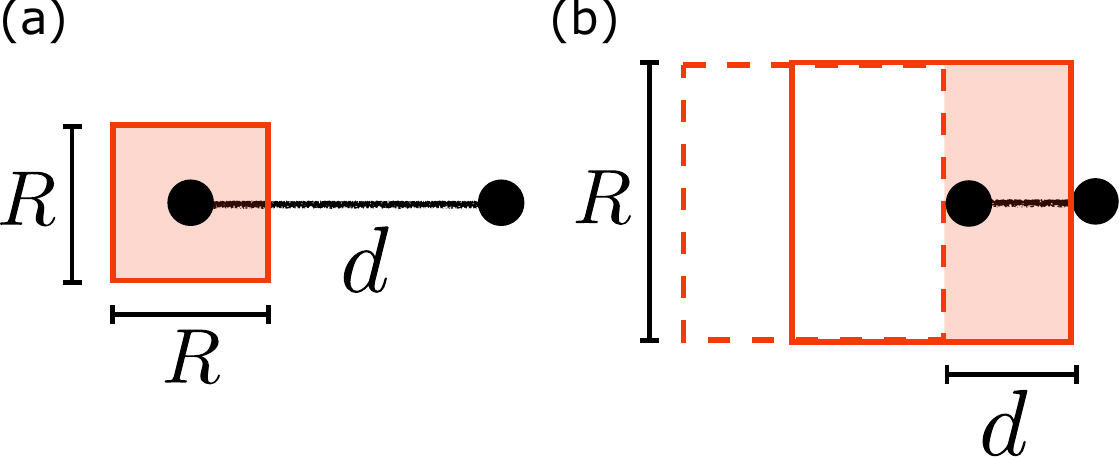}
\caption{Counting the number of squares of side length $R$ that enclose a single anyon in a $d$-separated pair. (a) If $R < d$, the square can be translated by $R$ units in both directions while still enclosing a single anyon. (b) If $R > d$, the square can be translated by $R$ units vertically but only by $d$ horizontally without crossing the other anyon.
}
\label{fig:area}
\end{figure}

We consider two magnetic anyons separated by a vector $\v{d}$ on the lattice and calculate their energy gap above the ground state, $\Delta E(\v{d})$. This energy is set by the number of loop stabilizers $\mathcal B_{\mathfrak s}$ that are flipped from their ground state value of $+1$. Note that only loops containing a single anyon are violated, as shown in Fig.~\ref{fig:area}. The energy of the two-anyon excitation therefore decomposes into a sum over all lattice loops that enclose a single anyon:
\begin{equation}
    \Delta E(\v{d})
    = 4 \sum_{\mathfrak s \in \mathcal{S}} \frac{\delta_{n_\mathfrak{s},1}}{\ell(\mathfrak{s})^\alpha},
    \label{eq:deltaE_general_sq}
\end{equation}
where $n_{\mathfrak s}$ is the number of anyons enclosed by the square loop $\mathfrak s$, and the factor of 4 comes from the contribution from the two anyons.
It is helpful to reorganize this expression into a sum over the loop perimeter $\ell$. For square loops, we find
\begin{equation}\label{eq:deltaE_square}
    \Delta E(\v{d}) = 4 \sum_{\ell} \frac{N_\text{enc}(\ell, \v{d})}{\ell^\alpha},
\end{equation} 
where $N_\text{enc}(\ell, \v{d})$ counts the number of ways a square of perimeter $\ell = 4R$ can be translated while enclosing one anyon but not the other, as illustrated in Fig.~\ref{fig:area}.

For concreteness, let us take the anyons to be separated horizontally by $\v{d} = d\hat{\v{x}}$, as in Fig.~\ref{fig:area}. The sum can then be decomposed into two terms. When $R \leq d$,
a square containing one anyon can be translated by $R$ units in both directions without enclosing the other anyon, giving $N_\text{enc} = R^2$. For $R> d$, the square can still be translated by $R$ units vertically, but only by $d$ units horizontally, giving  $N_\text{enc} = R d$.
Inserting these expressions for $N_\text{enc}(\ell, d)$ into Eq.~\eqref{eq:deltaE_square}, we finally obtain 
\begin{equation}
\label{eq:deltaE_square_twosum}
    \Delta E(d) = 4\sum_{R = 1}^{d} \frac{R^2}{(4R)^\alpha } + 4 \sum_{R = d+1}^L \frac{R d}{ (4R)^{\alpha }  }.
\end{equation}
For a generic anyon separation $\v{d}$ yields 
the same large-$d$ scaling for $\Delta E(\v d)$, up to an angle-dependent prefactor.

In the following subsections, we examine the asymptotic behavior of $\Delta E(d)$ in two regimes, $2<\alpha\leq3$ and $1<\alpha\leq2$. In the former regime, $\Delta E(d)$ grows algebraically or logarithmically with $d$. Consequently, individual anyons are confined, and   
a tightly bound anyon pair, or {\it dipole}, becomes a fundamental finite-energy excitation. In the latter regime, even a tightly bound dipole has an excitation energy that diverges with system size. The simplest finite-energy excitations are instead four-anyon square clusters, or anyon {\it quadrupoles}.

\subsection{Anyon confinement: $2 < \alpha \leq 3$}

For $\alpha > 2$, the large-$d$ behavior of \eqref{eq:deltaE_square_twosum} is given by:
\begin{equation}
    \label{eq:deltaE_square_scaling}
    \Delta E(d) \sim
    \begin{cases}
        d^{3-\alpha}, & 2 < \alpha < 3 , \\
        \ln d, & \alpha = 3, \\
        \text{const.}, & \alpha > 3.
    \end{cases}
\end{equation}
In the $2 < \alpha < 3$ regime, the loop interactions induce sublinear power-law confinement of anyons. At $\alpha=3$, there is logarithmic confinement.
On the other hand, the anyon separation energy is bounded for $\alpha > 3$, so the anyons are deconfined as in the ordinary toric code. 

The confining interaction for $2 < \alpha \leq 3$ can be used to estimate the probability that two isolated anyons are separated by a distance $d$: 
\begin{equation}\label{eq:prob_d}
    P(d) \sim d\, e^{-\beta \Delta E(d)},
\end{equation}
where $\beta = 1/T$ is the inverse temperature, and the entropic prefactor $d$ counts the number of possible orientations of anyon pairs of size $d$ at a fixed position. The first moment of this distribution gives the typical length scale $\xi$ for anyon separation. In the confining regimes, this length scale is
\begin{equation}\label{eq:anyon_xi}
    \xi \sim \begin{cases}
                \int_1^L x^2 e^{-\beta c x^{3-\alpha}} \,dx, & \Delta E(d) \sim d^{3-\alpha}, \\
                \int_1^L x^{2-c\beta}\,{dx}, & \Delta E(d) \sim \ln d,
    \end{cases}
\end{equation}
where $c$ is a constant that depends on the prefactor of $\Delta E(d)$.
For power-law confinement, $\xi$ is always finite, while for logarithmic confinement, $\xi$ is finite for $\beta > 3/c$.
Eq.~\eqref{eq:anyon_xi} provides a useful picture of a system at low temperatures, where anyons appear only rarely as tightly bound pairs of linear size $\xi$. 

At sufficiently low temperatures, the density of anyon dipoles is exponentially suppressed, motivating a dilute gas picture of tightly bound dipoles. Although these dipoles interact through the long-range loop terms, their mutual interaction can be shown to vanish at large separations. 
The dilute dipole gas thus provides a useful starting point for describing the low-temperature physics in the regime $2< \alpha \leq3$, although larger bound anyon clusters may also contribute.\footnote{In fact, an elementary square-shaped four-anyon cluster (anyon quadrupole) can have a lower excitation energy than a nearest-neighbor dipole. Explicitly, their excitation energies are $\Delta_2=4^{1-\alpha}\zeta(\alpha-1)$ and $\Delta_4=2\cdot4^{1-\alpha}\zeta(\alpha)$, where $\zeta$ is the Riemann zeta function, so $\Delta_4<\Delta_2$ for $2<\alpha<\alpha^*\approx2.479$, where $\alpha^*$ satisfies $2\zeta(\alpha^*)=\zeta(\alpha^*-1)$.
}

\subsection{Anyon super-confinement: $1 < \alpha \leq 2$}\label{subsec:ultra}

For $1 < \alpha \leq 2$, the leading term of \eqref{eq:deltaE_square_twosum} at $1 \ll d \ll L$ is 
\begin{equation}
    \label{eq:deltaE_square_scaling_lowalpha}
    \Delta E(d) \sim
    \begin{cases}
        d L^{2-\alpha}, & 1 < \alpha < 2 , \\
        d \ln{[L/d]} & \alpha = 2, \\
    \end{cases}
\end{equation}
resulting in a diverging energy cost for an anyon dipole as $L\to \infty$.
Thus, anyon dipoles are not finite-energy excitations, although $H_\text{sq}$ remains a thermodynamically well-defined many-body system for $\alpha >1$. 
Nevertheless, a pair of anyon dipoles can form a finite-energy bound state. 

To see this, consider four anyons arranged in a square of side length $d$. A square loop contributes to the excitation energy only if it encloses exactly one of the four anyons (enclosing three is impossible). Repeating the preceding loop-counting argument, we obtain
\begin{equation}\label{eq:four_anyon_energy}
    \Delta E_{\text{quad}} (d) = 8 \sum_{R=1}^d \frac{R^2}{(4R)^\alpha} + 8 \sum_{R=d+1}^L \frac{d^2}{(4R)^{\alpha}}.
\end{equation}
In the $L\to \infty$ limit, this expression scales at large $d$ as
\begin{equation}
    \label{eq:deltaE_quadrupole}
    \Delta E_\text{quad}(d) \sim d^{3-\alpha}.
\end{equation}
Thus, unlike an anyon dipole, a compact anyon quadrupole has finite excitation energy. 
At sufficiently low temperatures, this motivates a dilute gas picture of tightly bound anyon quadrupoles.

The energy of a four-anyon cluster depends sensitively on its geometry. 
For example, although four anyons at the corners of a rectangle have finite energy, four anyons occupying consecutive sites along a single row or column have an energy that diverges with system size. More generally, only configurations with an even number of anyons on each row and column of the lattice can have finite energy. These row- and column-neutrality conditions give rise to an emergent $\mathbb{Z}_2$ subsystem symmetry: anyon parity is separately conserved on every row and column.

Such subsystem symmetries are reminiscent of fracton systems, where they restrict mobility and lead to unusual dynamics~\cite{Nandkishore2019Mar, Pretko2020Mar}.
In the super-confining regime of $H_\text{sq}$, analogous mobility constraints emerge energetically. 
In particular, moving an individual anyon changes the parity of at least one row or column and therefore requires an energy that diverges with system size. Consequently, individual anyons become
immobile under finite-energy local {thermal} dynamics. 
We return to these constrained dynamics in Sec.~\ref{sec:memory}, where we discuss their role in protecting quantum information at finite temperatures.

\section{Anyon energetics at $T > 0$}\label{sec:energetics_finiteT}
The confinement derived in Sec.~\ref{sec:anyon_energetics} concerns isolated anyon configurations above the ground state. At finite temperatures, however, a background of thermally excited anyons can modify the energy cost of separating an additional anyon pair. We now examine this effect at low but finite temperatures. For $2< \alpha \leq3$, we show that confinement is screened by the thermal background beyond a large but finite length scale. For $1<\alpha\leq2$, on the other hand, we argue that anyon confinement persists at sufficiently low temperatures.

\subsection{Partial anyon confinement: $2 < \alpha \leq 3$}
\begin{figure}[t!]
\centering
\includegraphics[width=\linewidth]{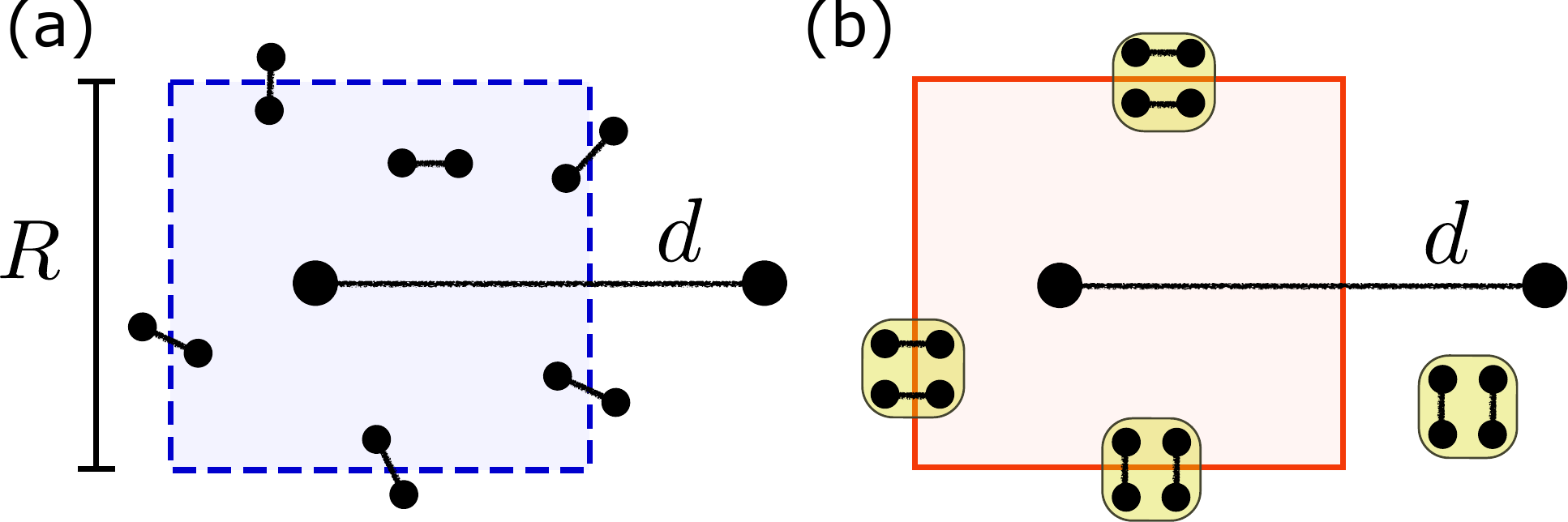}
\caption{
Thermal screening of the confining potential. (a) For $2<\alpha \leq 3$, a thermal background of anyon dipoles screens the confining interaction beyond a length scale $L_{\rm conf}$ \eqref{eq:Lconf}. 
(b) For $1<\alpha \leq 2$, the finite-energy excitations are anyon quadrupoles, which do not screen the confining potential, allowing confinement to persist at arbitrary scales.
} 
\label{fig:instability}
\end{figure}

Consider a gas of small anyon dipoles, randomly distributed throughout the system with a density $n_\text{pair}$. At low temperatures, this provides a simplified description of the thermal background in the long-range toric code. We now calculate the energy cost of separating two anyons a distance $d$ in this background, as depicted in Fig.~\ref{fig:instability}(a).

Without the background, this energy is a sum over the loops that enclose exactly one of the two anyons, as in Eq.~\eqref{eq:deltaE_square}. 
However, when a loop is intersected by a background anyon pair, its contribution to the energy changes sign. These fluctuations in sign reduce the mean energy cost associated with large loops, and Eq.~\eqref{eq:deltaE_square} is modified to
\begin{equation}\label{eq:background_energy}
    \Delta E (d) = 4 \sum_{R} \frac{N_\text{enc}(R, d)}{(4R)^\alpha} \expval{\mathcal B_{\mathfrak s(R)}},
\end{equation}
where $\expval{\mathcal B_{\mathfrak s(R)}}$ is the 
thermal expectation value of a loop stabilizer of side length $R$.\footnote{In the absence of the thermal background, all loops enclosing an anyon have the same sign, and the zero-temperature expectation value $\expval{\mathcal B_{\mathfrak s(R)}} = +1$ is implicitly present in Eq.~\eqref{eq:deltaE_square}.}

Assuming that the background pairs are independent and tightly bound (with separation $\xi = 1$): 
\begin{equation}
    \label{eq:typical_loop}
    \expval{\mathcal B_{\mathfrak s(R)}} \approx  (1 - n_\text{pair})^{4R} \approx  e^{-4 n_\text{pair} R},
\end{equation}
where the first approximation follows from the probability that each link of the loop is crossed by an anyon pair, and the second approximation holds for $n_\text{pair}\ll 1$. The interactions $\mathcal B_{\mathfrak s(R)}$ are Wilson loops, and Eq.~\eqref{eq:typical_loop} is the perimeter law that also characterizes the $T=0$ topological phase. In Appendix~\ref{app:wilson}, we provide a detailed treatment of contractible Wilson loops in the LRTC.

The energy of large loops in Eq.~\eqref{eq:background_energy} is exponentially suppressed with their perimeter and does not contribute to the sum, giving a crossover length scale: 
\begin{equation}\label{eq:Lconf}
    L_\text{conf} = \frac{1}{4n_\text{pair}} \sim e^{\beta \Delta_2},
\end{equation}
where $\Delta _2$ is the energy of the tightly bound anyon pair. If $d \ll L_\text{conf}$, then the confining potential  grows with separation $d$. Once $d \gtrsim L_\text{conf}$, however, the energy becomes approximately constant: 
\begin{equation}
    \label{eq:deltaE_square_scaling_background}
    \Delta E(d > L_\text{conf}) \sim
    \begin{cases}
        L_\text{conf}^{3-\alpha}, & 2 < \alpha < 3 , \\
        \ln{L_\text{conf}}, & \alpha = 3. \\
    \end{cases}
\end{equation}
The confinement of anyons is therefore screened beyond an $L$-independent length scale. 

\begin{figure*}[t!]
\centering
\includegraphics[width=0.83\textwidth]{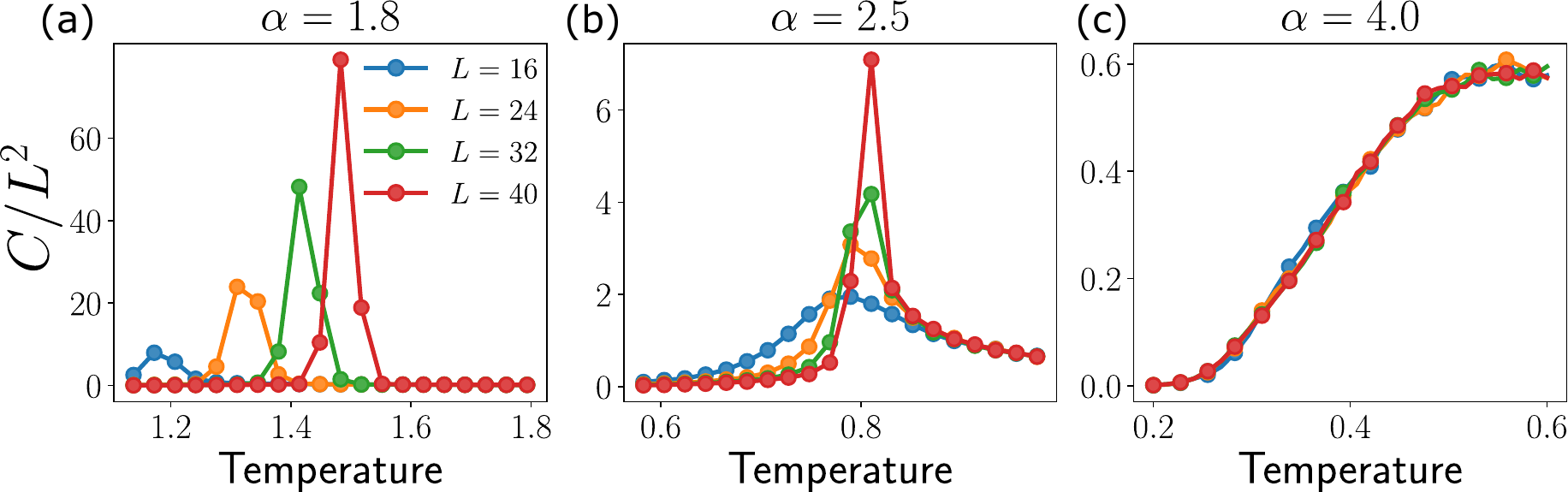}
\caption{Specific heat $C/L^2$ as a function of temperature in the square-loop model. (a) For $\alpha = 1.8$, $C/L^2$ has a peak that grows and drifts to higher temperatures as system size $L$ increases. (b) For $\alpha = 2.5$, $C/L^2$ has a peak that grows with $L$ around $T^* \approx 0.8$. These peaks are consistent with a thermodynamic transition in (a) and a sharp finite-size crossover in (b). (c) In the deconfined regime of $\alpha = 4$, the specific heat shows no appreciable dependence on $L$.}
\label{fig:heat}
\end{figure*}

A more careful treatment of the thermal background shows that the perimeter law of the Wilson loop is replaced by a power-law decay at asymptotically large $R$. More precisely, in Appendix~\ref{app:wilson} we establish that
$\expval{\mathcal B_{\mathfrak s(R)}} \sim {\beta }/ {R^\alpha}.$ as $R\rightarrow \infty$.
This unusual scaling evades bounds on Wilson loops in local gauge theories~\cite{Simon1982}. 
Inserting this power-law scaling 
into 
Eq.~\eqref{eq:background_energy} still gives a bounded anyon separation energy $\Delta E(d) \sim \Delta E^{\infty} - d^{3 - 2\alpha}$ as $d\to \infty$, where $\Delta E^{\infty}$ is a finite constant. The conclusions therefore remain the same: confinement breaks down at a large but finite length scale.

\subsection{Anyon confinement: $1 < \alpha \leq 2$}
For $2<\alpha \leq 3$, confinement is screened by background anyon dipoles. When $1 < \alpha \leq 2$, however, dipoles are not finite energy excitations. Instead, tightly bound anyon quadrupoles, which are composed of two adjacent dipoles, are the fundamental finite-energy excitations.
At low temperatures, the thermal background can then be approximated by a dilute gas of such quadrupoles, as depicted in Fig.~\ref{fig:instability}(b).

The expectation value of a large square Wilson loop in this background determines whether anyon confinement is screened at large distances. At $ T> 0$, a loop generically intersects many background quadrupoles. However, a quadrupole crossing a straight segment of the loop does not change its sign, since it inserts an even number of anyons inside the loop (see Fig.~\ref{fig:instability}(b)). Only quadrupoles that intersect a square loop at a corner can flip its sign. Assuming a dilute gas of independent elementary quadrupoles, the analogue of Eq.~\eqref{eq:typical_loop} is then
\begin{equation}\label{eq:wilson_constant}
\expval{\mathcal B_{\mathfrak s(R)}} \approx (1 - 2n_\text{quad})^4 \approx e^{-8 n_\text{quad}},
\end{equation} 
where $n_\text{quad}$ is the density of quadrupoles. Thus, the thermal background alters the Wilson loop expectation value by a constant factor that is independent of the loop size $R$.

Unlike the $2<\alpha\leq3$ regime, the thermal background of quadrupoles therefore does not screen confinement at large distances. Consequently, the anyon separation energy retains the same asymptotic scaling as its zero-temperature counterpart in Sec.~\ref{subsec:ultra}. Anyon confinement should then persist at sufficiently low temperatures for $1<\alpha\leq2$, providing a mechanism for finite-temperature topological order in the thermodynamic limit.

\section{Thermodynamic transition}\label{sec:transition}

For $1<\alpha\leq2$, our analysis of finite-temperature anyon confinement suggests that $H_\text{sq}$ may support a low-temperature topologically ordered phase. Such a phase would be separated from the high-temperature trivial phase by a thermodynamic transition. We now investigate this possibility with Monte Carlo simulations of the long-range toric code, sampling equilibrium configurations of anyons at various temperatures.

The Hamiltonian $H_\text{sq}$ consists of mutually commuting stabilizers, each involving exclusively either Pauli $X$ or Pauli $Z$ operators. Such a structure defines a Calderbank-Shor-Steane (CSS) stabilizer code~\cite{Calderbank1996,Steane1996}. The partition function of a CSS code factorizes into two independent classical partition functions, one for the $X$ sector and one for the $Z$ sector~\cite{Weinstein2019}. Since the two sectors have identical thermodynamics, it suffices to perform classical Monte Carlo simulations on only the  $Z$ (magnetic) sector. Focusing on $\alpha=1.8$, $2.5$, and $4$, we sample $10^4$ equilibrated configurations for 30 different temperatures. Further details of the Monte Carlo simulations are provided in Appendix~\ref{app:montecarlo}.

To detect a possible transition without needing access to a specific order parameter, we calculate the heat capacity:
\begin{equation}
    \label{eq:heat_capacity}
    C = \beta^2 \left(\expval{E^2}_\beta - \expval{E}_\beta^2\right),
\end{equation}
where $E$ is the classical energy associated with the magnetic sector of $H_\text{sq}$, and $\langle{\,}\rangle_\beta$ indicates a thermal expectation value at temperature $T = 1/\beta$. A thermodynamic phase transition can manifest as a nonanalyticity in the specific heat $C/L^2$ in the thermodynamic limit~\cite{Kardar2007}. 

The results of our simulations are shown in Fig.~\ref{fig:heat}, where the specific heat is plotted as a function of temperature for a series of system sizes $L$. 
For $\alpha = 1.8$ and $2.5$, the specific heat has a peak that increases with system size.
For $\alpha = 1.8$, this is consistent with a thermodynamic transition, although the peak drifts considerably toward higher temperatures as $L$ increases.
For $\alpha=2.5$, the peak occurs around $T^* \approx0.8$. 
According to the arguments of Sec.~\ref{sec:energetics_finiteT}, this peak should be interpreted as a sharp finite-size crossover rather than a thermodynamic transition into a topologically ordered phase.
For $\alpha = 4$, no size-dependent peak is observed in the specific heat, consistent with the fact that anyons are deconfined.

Our analytical arguments and numerical simulations suggest that $H_\text{sq}$ has a thermodynamic phase transition in the confining regime $1<\alpha\leq2$ and exhibits a sharp crossover in the partially confining regime $2<\alpha \leq 3$. We leave the precise characterization of these behaviors to future work and focus on the low-temperature properties in the remaining sections.

\section{Finite-temperature quantum memory}\label{sec:memory}

A topologically ordered system can be characterized by the quantum information encoded in its ground state manifold. For example, the toric code on a torus encodes two logical qubits. 
At finite temperatures, the stability of this information under thermal noise provides an operational definition of finite-temperature topological order. A many-body system that protects quantum information for a time that diverges with system size, without active error correction, is referred to as a passive, or self-correcting, quantum memory. In this section, we investigate the memory lifetime of the long-range toric code in the fully and partially confining regimes.

\subsection{Memory lifetime}

A precise definition of the quantum memory lifetime requires three inputs: an encoding protocol to store information, a model of thermal dynamics, and a decoding protocol to retrieve information. We encode information by preparing a ground state of $H_\text{sq}$, $\rho_0=\ket{\psi_\text{tc}}\bra{\psi_\text{tc}}$.
Assuming that our system is weakly coupled to an external bath, we describe its thermal dynamics by a Lindblad equation:
\begin{equation}\label{eq:master_lindblad}
    \frac{d\rho(t)}{dt} = -i[H_\text{sq}, \rho(t)] + \mathcal{L}[\rho(t)], 
\end{equation}
where the Lindbladian $\mathcal{L}$ governs how the system and bath exchange energy. 
We assume local system-bath couplings, with transition rates satisfying detailed balance so that the Gibbs state $\rho_\beta \propto e^{-\beta H_\text{sq}}$ is a steady state of the dynamics~\cite{Alicki2009, Alicki2012, Bravyi2013haah}. 

For a CSS code like $H_\text{sq}$, it is conventional to model the coupling to a thermal bath using local Pauli $X$ and $Z$ errors.
The dynamics in Eq.~\eqref{eq:master_lindblad} then reduces to classical Gibbs sampling of the $X$ and $Z$ sectors of the Hamiltonian, which can be simulated using Metropolis updates~\cite{Siva2017}.
For instance, an $X$ error at site $i$ flips the corresponding $Z$ spin with a rate
\begin{equation}\label{eq:metropolis}
    p(Z_i \mapsto - Z_i) = \min(1, e^{-\beta \Delta E}),
\end{equation}
where $\Delta E$ is the classical energy difference due to the proposed spin flip, and the Boltzmann factor ensures that the dynamics satisfies detailed balance.

After the system evolves for some time at finite temperature, it becomes 
populated with anyon defects. To extract the information stored in the memory, these defects must be cleaned up by a decoder $\mathcal{D}$. Successful decoding returns the system to its original logical state $\mathcal{D}[\rho(t)] = \rho_0$, whereas a decoding failure produces a logical error.
For the long-range toric code, we use a minimum-weight perfect-matching (MWPM) decoder, which annihilates anyons in pairs in a way that minimizes the total length of the correction strings~\cite{Dennis2002}.

Given our encoding map, thermal dynamics, and a decoding map, the memory lifetime is defined as:
\begin{equation}\label{eq:tmem}
    t_\text{mem} = \min \{t: \, \Vert \mathcal{D}[\rho(t)] - \rho(0) \Vert_1 > \epsilon \},
\end{equation}
where $\Vert \, \Vert_1$ is the trace norm and $\epsilon$ is a fixed positive constant. Eq.~\eqref{eq:tmem} represents the first time that the distance between the encoded state and the decoded state exceeds a finite threshold $\epsilon$.
In our numerical simulations, we instead take $t_\text{mem}$ to be the first time at which a MWPM decoder maps $\rho(t)$ to a different ground state from $\rho_0$. A self-correcting quantum memory is a many-body system of linear size $L$ such that
\begin{equation}\label{eq:quantum_memory_def}
    \lim_{L\rightarrow \infty} t_\text{mem} = \infty.
\end{equation}

The most famous example of a self-correcting quantum memory is the 4D toric code~\cite{Dennis2002}. In lower dimensions, however, quantum memories are difficult to construct. For example, local 2D and 3D models described by a topological quantum field theory cannot host a self-correcting quantum memory because of the proliferation of pointlike  defects~\cite{Bravyi2009nogo,Yoshida2011}. 
In 3D, models with fractal excitations like the Haah code~\cite{Haah2011Apr} (see also ~\cite{Michnicki2014Sep, Williamson2024Nov, Baspin2025Sep,Williamson2026Aug}) were designed to circumvent this obstruction. However, $t_\text{mem}$ was later shown to grow only up to a finite, temperature-dependent length scale, making the memory only \emph{partially self-correcting}~\cite{Bravyi2013haah, Siva2017}.

Thus far, known constructions of fully or partially self-correcting quantum memories with local interactions are restricted to 3D and above~\cite{Balasubramanian2026}. Our long-range toric code circumvents the 2D locality obstruction through long-range loop interactions.
In the following sections, we argue that $H_\text{sq}$ exhibits partial self-correction for $2<\alpha\leq3$ and provide evidence for full self-correction for $1<\alpha\leq2$.

\subsection{Partially self-correcting regime: 
$2< \alpha \leq 3$}

In the long-range toric code, the memory fails when $\rho(t)$ can no longer be reliably decoded back to the initial toric code ground state. 
This could occur when two anyons separate by an $\mathcal{O}(L)$ distance, potentially producing a noncontractible error string.
Heuristically, the probability of this process is given by the Boltzmann factor $e^{-\beta \Delta E}$, where $\Delta E$ is the minimum energy cost of separating the anyons. The associated memory time,
\begin{equation}\label{eq:arrhenius}
    t_\text{mem} \sim e^{\beta \Delta E},
\end{equation}
is given by the Arrhenius law. Much of the literature on quantum memories has focused on finding systems with energy barriers $\Delta E$ that grow with $L$~\cite{Brown2016Nov}. 

However, the Arrhenius estimate ignores entropic factors: the probability that the memory fails depends on the number of possible failure processes, $\Omega = e^{\Delta S}$. As a result, a true quantum memory actually requires a diverging \emph{free energy} barrier:
\begin{equation}\label{eq:free_energy_barrier}
    t_\text{mem} \sim e^{\beta \Delta F},
\end{equation}
with $\Delta F = \Delta E - T \Delta S$ being the minimum free energy required to incur a logical error. Aside from a few exceptions~\cite{Hong2025ldpc, Placke2024ldpc}, determining the precise entropic factor is intractable for a many-body quantum system. Nevertheless, we can estimate $t_\text{mem}$ heuristically using the energy barrier required to separate two anyons by an $\mathcal{O}(L)$ distance in a thermal background.

For system sizes $L < L_\text{conf}$, anyons remain confined at finite temperatures, and the energy barrier grows with system size. Using Eq.~\eqref{eq:deltaE_square_scaling}, we have
\begin{equation}
    \label{eq:longtoric_arrhenius}
    t_\text{mem} (L < L_\text{conf}) \sim
    \begin{cases}
        \exp [\beta c L^{3-\alpha}], & 2 < \alpha < 3 , \\
        L^{c\beta}, & \alpha = 3, \\
        \text{const.}, & \alpha > 3,
    \end{cases}
\end{equation}
where $c$ is a constant. However, the initial growth of the memory time with $L$ is only a transient behavior. Once $L > L_\text{conf}$, the thermal background screens the confinement, and the energy barrier no longer grows with system size. As a result, $t_\text{mem}$ stops growing with $L$, and the quantum memory is merely partially self-correcting.

Nevertheless, quantum information can still be reliably stored for extremely long times. 
Using the energy barrier in Eq.~\eqref{eq:deltaE_square_scaling_background} and $L_\text{conf} \sim e^{\beta \Delta_2}$, where $\Delta_2$ is
the energy of a tightly bound anyon pair,
the memory time saturates to
\begin{equation}
    t_\text{mem} (L > L_\text{conf}) \sim 
        \begin{cases}
        \exp\left[c \beta \exp\left[c'\beta \Delta_2 \right] \right], & 2 < \alpha < 3, \\
         \exp\left[c \beta^2 \Delta_2 \right], & \alpha = 3,\\
    \end{cases}
\end{equation}
where $c$ and $c'$ are constants. Thus, beyond a system size that is exponentially large in $\beta$, the memory time is super-exponentially large in $\beta$. 
Remarkably, $H_\text{sq}$ provides the first example of partial self-correction below three dimensions, with the 
memory lifetime exhibiting the strongest known temperature dependence~\cite{Williamson2026Aug}.
Next, we turn to the regime $1 < \alpha \leq 2$, where the lack of screening promotes partial self-correction to a fully self-correcting quantum memory.  

\begin{figure}[t!]
\centering
\includegraphics[width=0.5\linewidth]{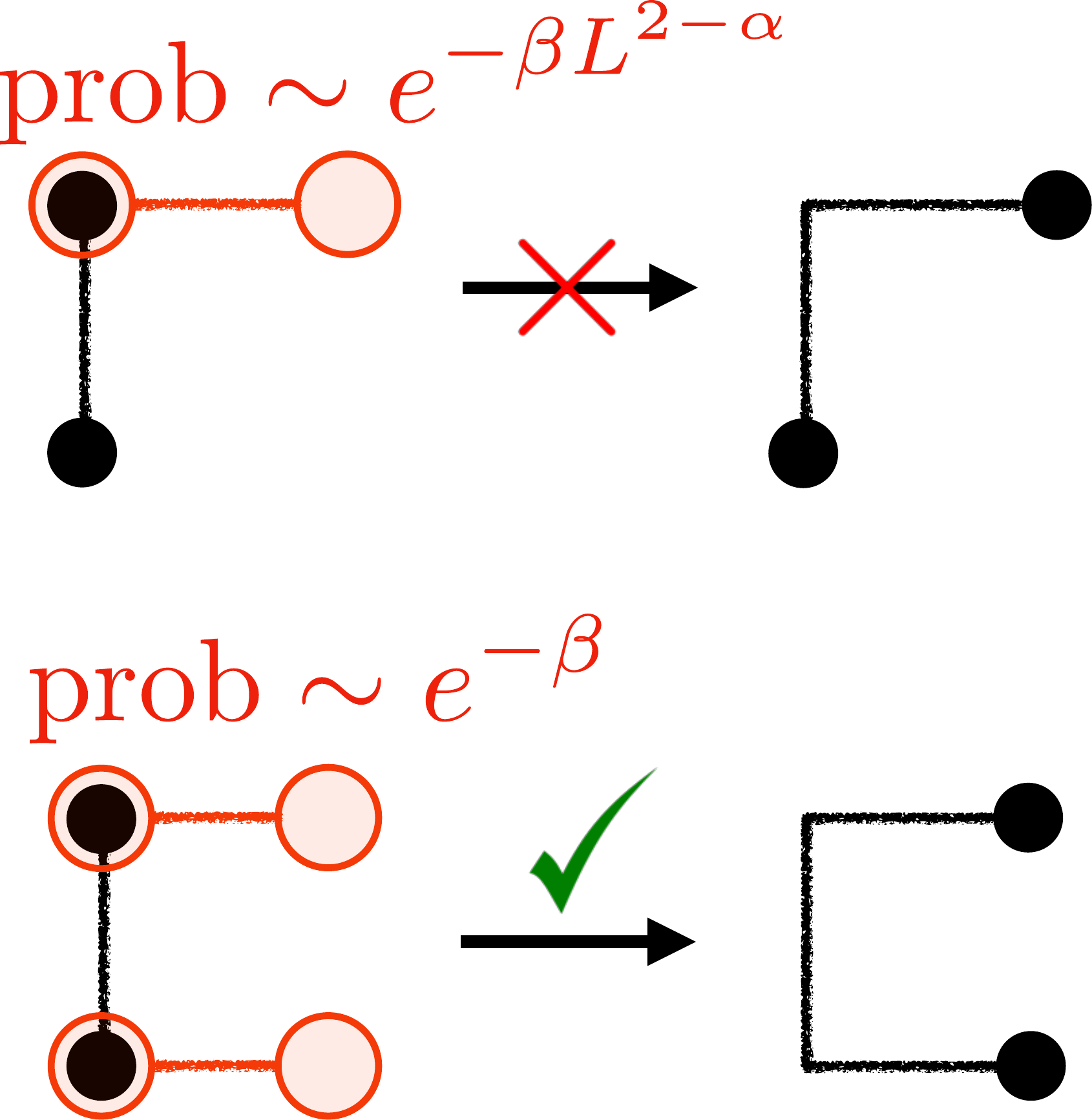}
\caption{Top: An individual anyon moves by absorbing 
an anyon dipole from the thermal bath. This process has vanishing probability as $L\rightarrow \infty$. 
Bottom: An anyon dipole moves by absorbing an anyon quadrupole from the bath. This process occurs with finite probability.
} 
\label{fig:fracton}
\end{figure}

\subsection{Fully self-correcting regime: $1 < \alpha \leq 2$} 

In the $1 < \alpha \leq 2$ regime, we have argued in Sec.~\ref{subsec:ultra} that anyon confinement persists to low but finite temperatures.
The energy barrier in Eq.~\eqref{eq:deltaE_square_scaling_lowalpha} therefore remains valid at finite temperature, yielding
\begin{equation}\label{eq:tmem_ultra}
    t_\text{mem} \sim 
        \exp\left[\beta L^{3-\alpha} \right]  \quad (1 < \alpha \leq 2).
\end{equation}
This estimate of $t_\text{mem}$ diverges as $L\rightarrow \infty$, suggesting that $H_\text{sq}$ is a genuine self-correcting quantum memory in this regime. 

\begin{figure*}[t!]
\centering
\includegraphics[width=0.85\textwidth]{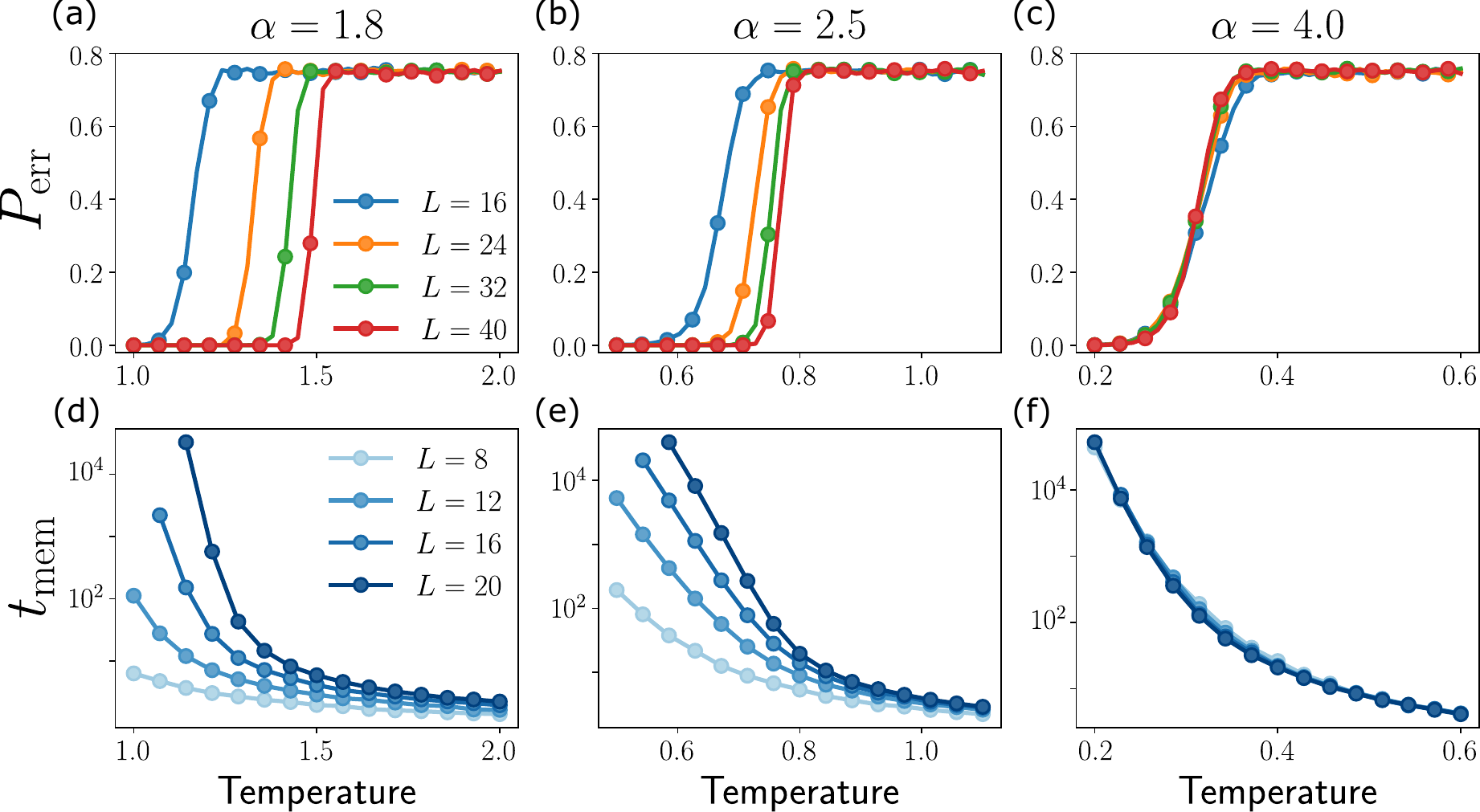}
\caption{Numerical simulations of quantum memory in $H_\text{sq}$. Top: Logical error probability $P_\text{err}$ after decoding as a function of temperature, for a range of system sizes $L$. Simulations begin in a ground state and run for $200$ Monte Carlo steps before decoding. For (a) $\alpha = 1.8$ and (b) $\alpha=2.5$, the probability decreases with $L$ in the low temperature regime, while for (c) $\alpha = 4$, the probability is approximately $L$-independent.
Bottom: Average time $t_\text{mem}$ at which the memory fails. For (d) $\alpha = 1.8$ and (e) $\alpha=2.5$, $t_\text{mem}$ grows with $L$ in the low-temperature regime, suggesting a quantum memory, while for (f) $\alpha = 4$, $t_\text{mem}$ is approximately $L$-independent.
}
\label{fig:memory}
\end{figure*}

{The full self-correction can also be viewed as a consequence of kinetic constraints on anyon mobility.}
As outlined in Sec.~\ref{subsec:ultra}, finite-energy configurations require an even number of anyons on each row and column of the lattice. 
This constraint represents an emergent subsystem symmetry that restricts the motion of anyons. We now show how such mobility constraints suppress logical errors, potentially protecting
quantum information for arbitrarily long times.

Since anyon occupations commute with $H_\text{sq}$, the Hamiltonian time evolution in Eq.~\eqref{eq:master_lindblad} cannot move anyons, and their mobility comes entirely from interactions with the thermal bath. These interactions create and annihilate local clusters of anyons, and existing anyons then move by fusing with these clusters. A single anyon moves when {it absorbs an anyon dipole created by the bath},
as shown at the top of Fig.~\ref{fig:fracton}. Similarly, an anyon dipole moves by absorbing a quadrupole created by
the bath---depicted at the bottom of Fig.~\ref{fig:fracton}.
This mediated mobility is reminiscent of fractons, whose motion is constrained by dipole conservation~\cite{Nandkishore2019Mar, Pretko2020Mar}.

Dipole and quadrupole creation, however, occur at different rates. The energy of an isolated anyon dipole diverges as $L^{2-\alpha}$ {for $1<\alpha <2 $ and as $\ln L $ for $\alpha =2$}, giving a vanishingly small Boltzmann factor. 
In the thermodynamic limit, the creation of isolated dipoles is suppressed, and therefore individual anyons become immobile. 
Conversely, anyon quadrupoles have an $\mathcal{O}(1)$ energy and thus a finite Boltzmann factor. Since the bath creates quadrupoles at a finite rate, thermal dynamics move dipoles as a unit. 
During this process, anyon parity is independently conserved in each row and column, dynamically imposing the emergent subsystem symmetry.

Since isolated anyons must overcome an $\mathcal{O}(L^{2-\alpha})$ or $\mathcal{O}(\ln L)$ energy barrier to move, logical errors are severely suppressed. A logical error is topologically equivalent to moving an anyon around a noncontractible loop, a process that is forbidden by the mobility constraints. 
Moving a dipole around the torus, on the other hand, creates two noncontractible loops and thus incurs no logical error. As long as the kinetic constraints hold, the logical information remains preserved by thermal dynamics.

The anyon confinement and kinetic constraints provide strong evidence for a self-correcting quantum memory in the $1<\alpha \leq 2$ regime. However, a rigorous proof of a diverging memory time is challenging for a few reasons. First, the long-range interactions couple each spin to every other spin in the system, complicating a Peierls-style argument based on disconnected clusters of errors~\cite{Hong2025ldpc, Balasubramanian2026, Stahl2026Sep}. Second, the separation of two anyons by an $\mathcal{O}(L)$ distance is not the only ``dangerous'' configuration that leads to a logical error. 
{For example, the coordinated motion of many small anyon pairs can create a noncontractible loop without two nearby anyons ever separating by an $\mathcal{O}(L)$ distance.}
Lastly, the persistence of anyon confinement relies on
the nonzero Wilson loop asymptotics in Eq.~\eqref{eq:wilson_constant}, which 
is based on a heuristic description of the thermal state.

\subsection{Numerical evidence for the memory}

To corroborate the estimates of $t_\text{mem}$ in the previous sections, we numerically estimate the memory lifetime of $H_\text{sq}$. After preparing a toric code ground state, we simulate thermal dynamics using classical Monte Carlo and then apply MWPM decoding to the the final state (see Appendix~\ref{app:montecarlo} for further details). Whether the decoder recovers the original ground state or not determines if a logical error has occurred.

We employ two protocols. In the first protocol, we run thermal dynamics for a fixed number of time steps ($t=200$) and then decode. The probability that the decoder fails measures the logical-error probability after this evolution time. 
The results of this protocol are shown in Fig.~\ref{fig:memory}(a-c). For $\alpha = 1.8$ and $\alpha = 2.5$, the probability of a logical error decreases with system size at low temperatures, suggesting that the memory improves with increasing $L$. Although the $\alpha = 2.5$ regime is only partially self-correcting, its behavior resembles that of a fully self-correcting memory at accessible system sizes.
Furthermore, for $\alpha = 2.5$, the memory persists until $T^*\approx 0.8$, consistent with the specific heat peak in Fig.~\ref{fig:heat}(b). On the other hand, the failure probability for $\alpha = 4$, where anyons are always deconfined, is approximately independent of system size, consistent with the absence of self-correction.

The second protocol evolves a state until the first decoding failure,
thus extracting $t_\text{mem}$ directly. Because the memory time increases exponentially in inverse temperature, we are limited to smaller system sizes.
The memory time averaged over Monte Carlo runs is shown in Fig.~\ref{fig:memory}(d-f). For $\alpha = 1.8$ and $\alpha=2.5$, $t_\text{mem}$ appears to grow with system size at low temperatures, while it is $L$-independent for $\alpha = 4$. 
Again, both the fully and partially self-correcting regimes exhibit increasing memory protection
at accessible system sizes.
{In the partially self-correcting regime of $\alpha =2.5$, for instance, we estimate $L_\text{conf}\sim10^4$ at $T=0.5$, corresponding to a memory lifetime of $t_\text{mem}\sim10^{100}$.}

\begin{figure*}[t!]
\centering
\includegraphics[width=0.84\textwidth]{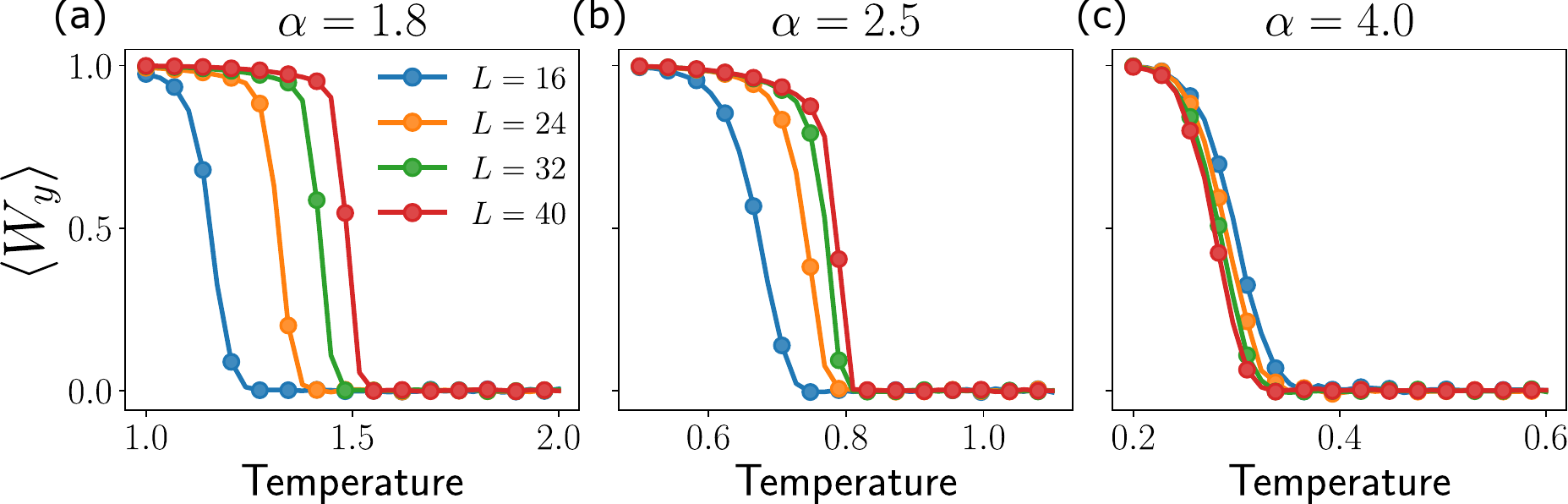}
\caption{
Expectation value of a non-contractible Wilson loop $W_y$ after 200 Monte Carlo steps.
All simulations are initialized in a ground state with $W_y = +1$. (a-b)
For $\alpha=1.8$ and $2.5$, $\langle W_y\rangle$ remains close to one at low temperatures, indicating that the initial logical information is largely preserved over the observation time. 
(c) For $\alpha=4$, $\langle W_y\rangle$ decays weakly with system size for any temperature.}
\label{fig:noncontract}
\end{figure*}

Another quantity related to the memory lifetime is the expectation value of a noncontractible Wilson loop $W_y = \prod_{j} Z_j$, where the product is over the links of a noncontractible cycle on the torus.
We initialize the Monte Carlo simulations in a toric code ground state with $W_y = + 1$. After running thermal dynamics, the Wilson loop expectation value indicates how close the finite-temperature state remains to a single logical sector~\cite{Bergamaschi2025}. In Fig.~\ref{fig:noncontract}, we show $\expval{W_y}$ calculated by averaging over $10^4$ trials after $200$ steps of initial thermal dynamics.
{For both the fully and partially self-correcting regimes, $\expval{W_y}$ is close to one at low temperatures, indicating that
thermal dynamics remains close to the original ground state. On the other hand, $\expval{W_y}$ weakly decays with system size for $\alpha=4$, consistent with no quantum memory. At the lowest temperatures, $\expval{W_y}$ nevertheless remains nonzero because the memory time exceeds the simulation time of $t=200$ steps (cf. Fig.~\ref{fig:memory}(f)).}

\section{Long-range entanglement}\label{sec:lre}

Thus far, we have characterized topological order dynamically through the lifetime of its quantum memory. We now turn to an equilibrium charaterization of topological order: long-range entanglement (LRE) of the Gibbs state. A pure state is said to be LRE if it cannot be prepared from a product state by a quasi-local unitary circuit~\cite{Chen2010Oct}. Here, quasi-local means that the support of the unitary gates and the circuit depth are both $\mathcal{O}(\text{polylog}\,L)$. States that can be prepared by such a circuit are instead called short-range entangled (SRE). In the mixed state setting, a density matrix $\rho$ is called LRE if it admits 
no decomposition
\begin{equation}\label{eq:mixed_lre}
    \rho = \sum_i p_i \ket{\phi_i}\bra{\phi_i},
\end{equation}
in which every $\ket{\phi_i}$ is SRE~\cite{Hastings2011}. 

If $H_\text{sq}$ is a finite-temperature quantum memory, it is likely that its Gibbs state $\rho_\beta \propto e^{-\beta H_\text{sq}}$ is long-range entangled. 
This is true for other mixed-state quantum memories, such as the 4D toric code at finite temperature~\cite{Dennis2002, Hastings2011} and the 2D dephased toric code~\cite{Fan2024, Lessa2025Mar}.
Moreover, LRE appears to be a weaker condition than quantum memory, as several LRE mixed states are only classical memories~\cite{Wang2025Jan, Zhou2025}. However, a precise relation between the two properties has not been established, 
so we present an independent argument that anyon confinement gives rise to LRE in $\rho_\beta$. 
Specifically, we argue that the low-temperature Gibbs state is genuinely LRE for $1<\alpha\leq2$, while the regime $2<\alpha < 3$ exhibits only finite-size signatures of LRE below $L_\text{conf}$.\footnote{We exclude $\alpha = 3$ from our argument, where anyon confinement is only logarithmic.}

Any mixed state has infinitely many decompositions into pure-state ensembles of the form in Eq.~\eqref{eq:mixed_lre}, making a proof of LRE challenging. 
Instead, we use a quasi-local channel,\footnote{A quasi-local channel is a channel whose Stinespring dilation is a quasi-local unitary circuit. In other words, the channel acts on a system by adding ancilla qubits to each site, evolving with a quasi-local unitary circuit, and then tracing out the ancillas.} 
a mixed state version of a quasi-local unitary circuit, to map $\rho_\beta$ to the zero-temperature Gibbs state:
\begin{equation}
    \label{eq:qlc_toric_general}
    \mathcal{P}_\text{QLC}[\rho_\beta] = \rho_{\infty},
\end{equation}
Here $\rho_\infty$ is  the maximally mixed state in the ground state subspace of the toric code, which is LRE~\cite{Li2025Jun}.
Since quasi-local channels map SRE states to SRE states and cannot generate long-range correlations~\cite{Sang2024, Zhou2025},
the existence of a channel satisfying Eq.~\eqref{eq:qlc_toric_general} 
implies that $\rho_\beta$ is LRE.

All that remains is to find a quasi-local channel that maps $\rho_\beta$ to the ground state subspace of the toric code. Such a channel is a decoder, cleaning up anyon defects in the thermal state. In Sec.~\ref{sec:memory}, we used a MWPM decoder to define the quantum memory lifetime. However, MWPM is not a quasi-local channel because it pairs up anyons globally across the system. Another decoder is therefore needed to establish LRE. To this end, we use the renormalization group (RG) decoder introduced in Ref.~\onlinecite{Sang2024}, which acts by iteratively coarse-graining anyon configurations. Since electric and magnetic defects are cleaned up independently, it suffices to consider a single type of defect.

\begin{figure}[t!]
\centering
\makebox[\linewidth][c]{%
    \includegraphics[width=0.63\linewidth]{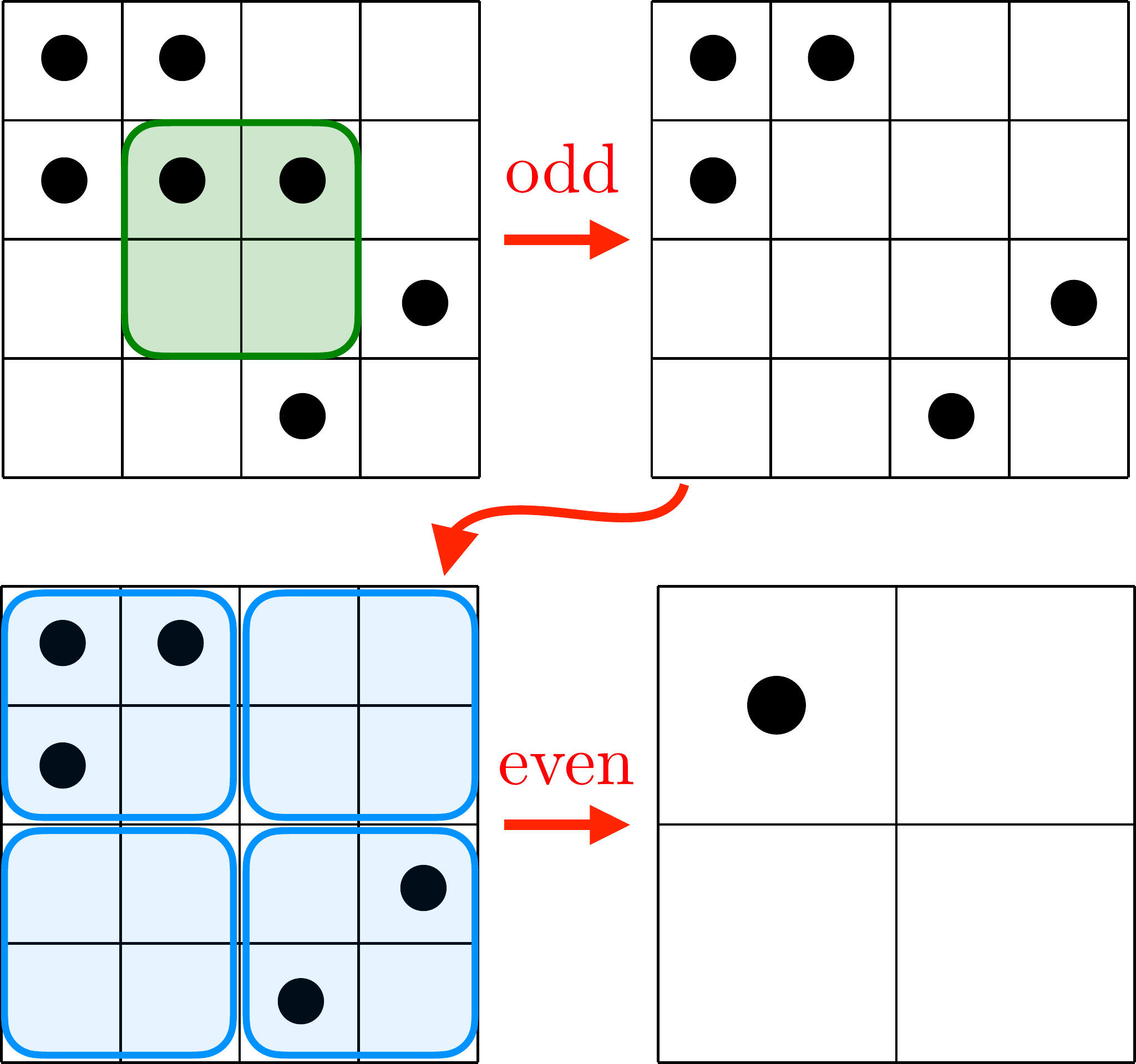}%
}
\caption{A sample step of the RG decoder introduced in Ref.~\onlinecite{Sang2024} on a subset of a larger lattice. First, anyons are annihilated inside odd squares (green), though only if they appear in even number. Next, even squares (blue) are coarse-grained while preserving their anyon parity.} 
\label{fig:rgstep}
\end{figure}

The decoder partitions an $L \times L$ lattice into two sets of $2 \times 2$ squares: even squares $S_\text{even}$ and odd squares $S_\text{odd}$, with the latter obtained by shifting the former horizontally and vertically by one lattice spacing. 
A single RG step, illustrated for a $4\times 4$ region in Fig.~\ref{fig:rgstep}, proceeds 
as follows. First, the decoder scans through every odd square. If an $S_\text{odd}$ contains an even number of anyons, the anyons are annihilated.
Next, each even square is coarse-grained into a single site with the same anyon parity. 
In other words, even squares containing one or three anyons are mapped to a site with a single anyon, while those containing zero, two, or four anyons are mapped to a site with no anyon. 

After coarse-graining, the lattice shrinks by a factor of two in each direction. More precisely, coarse-grained qubits are decoupled from the rest of the system by unitary gates through entanglement renormalization~\cite{Aguado2008Feb}. 
Iterating this procedure $k$ times defines an RG flow on the space of anyon configurations. Whether this flow decreases or increases the anyon density determines the success of the decoder. In addition, to be a \emph{quasi-local} channel, the decoder must clean up anyons before the scale of the RG grows too large. At the $k$-th step, the decoder acts on blocks of size $2^k$, meaning that $2^k \lesssim  \ln L$ for quasi-locality to hold.

To quantify the decoder's success, we bound the distance between the decoded state and the ground state subspace of the toric code, $V$. Let us denote $k$ applications of the decoder by
\begin{equation}
\rho^{(k)}
= \overbrace{\mathcal{P}_{\mathrm{RG}} \circ \cdots \circ \mathcal{P}_{\mathrm{RG}}}^{k\text{ times}}[\rho_\beta]
= \mathcal{P}_{\mathrm{RG}}^{\,k}[\rho_\beta].\end{equation} If the decoder is successful and quasi-local, $\rho^{(k)}$ tends to the ground state subspace in $k = \mathcal O (\ln \ln L)$ steps.
We are therefore interested in the overlap
\begin{equation}\label{eq:overlap}
F^{(k)} = \Tr(\rho^{(k)} \Pi_V),
\end{equation}
where $\Pi_V$ is a projector onto the ground state subspace of the toric code.
Physically, $F^{(k)}$ represents the probability that $\rho^{(k)}$ has exactly zero anyons.\footnote{Unlike a quantum memory experiment, where a particular toric code ground state is prepared and then evolved under thermal dynamics, the Gibbs state $\rho_\beta\propto e^{-\beta H_\text{sq}}$ is a mixture of all four logical sectors. The natural decoding target is therefore the entire ground state subspace $V$, rather than a particular ground state.} 

In the confining regime of $H_\text{sq}$, the low-temperature Gibbs state is approximately a dilute gas of tightly bound clusters of anyons. For $2 < \alpha \leq 3$, these clusters are anyon pairs, and remain bound only up to a finite system size. For $1 < \alpha \leq 2$, these clusters are anyon quadrupoles, and remain bound for $L\rightarrow \infty$. 
After $k$ steps, clusters of linear size $d \lesssim 2^k$ are annihilated by the RG decoder.
The probability that any anyons survive is bounded by
\begin{equation}
    \label{eq:prob_survival}
    1-F^{(k)} \lesssim n L^2 P(d > 2^k),
\end{equation}
where $n$ is the anyon density and $P(d > 2^k)$ is the probability that an anyon cluster is larger than $2^k$. Eq.~\eqref{eq:prob_survival} assumes that the decoder acts on each anyon cluster independently, which approximately holds at low temperatures, where clusters are dilute.

The probability in Eq.~\eqref{eq:prob_survival} can be estimated using the energetics of Sec.~\ref{sec:anyon_energetics}. Working in the power-law confining regime, Eq.~\eqref{eq:prob_d} gives
\begin{equation}\label{eq:prob_2k}
    P(d >2^k) \sim \int_{2^k}^\infty x e^{-\beta x^\gamma} dx,
\end{equation}
where $\gamma > 0$.\footnote{Strictly speaking, Eq.~\eqref{eq:prob_d} gives the size distribution of anyon pairs for $2 < \alpha < 3$, but a similar expression holds for quadrupoles when $1 < \alpha \leq 2$.}
At sufficiently low temperatures, this probability scales as $\exp\,[-\beta 2^{k\gamma}]$ up to polynomial prefactors. The decoder achieves an arbitrarily small failure probability, $1-F^{(k)}<\epsilon$, if
\begin{equation}\label{eq:rg_bound_scaling}
    n L^2 \exp\,[-\beta 2^{k\gamma}] \lesssim \epsilon.
\end{equation}
Inverting Eq.~\eqref{eq:rg_bound_scaling}, we finally obtain 
\begin{equation}
    k \gtrsim \frac{1}{\gamma \ln 2}\ln \left [\frac{1}{\beta} \ln \left(\frac{n L^2}{\epsilon}\right) \right].
\end{equation}
Since the RG decoder acts on regions of maximum size $2^k = \mathcal{O}(\text{polylog}\ L)$, it is a quasi-local channel. We have thus constructed a quasi-local channel that, for sufficiently low temperatures, maps the Gibbs states of $H_\text{sq}$ to the ground state subspace of the toric code. As a result, the Gibbs states must be long-range entangled.

\begin{figure}[t!]
\centering
\makebox[\linewidth][c]{%
    \includegraphics[width=\linewidth]{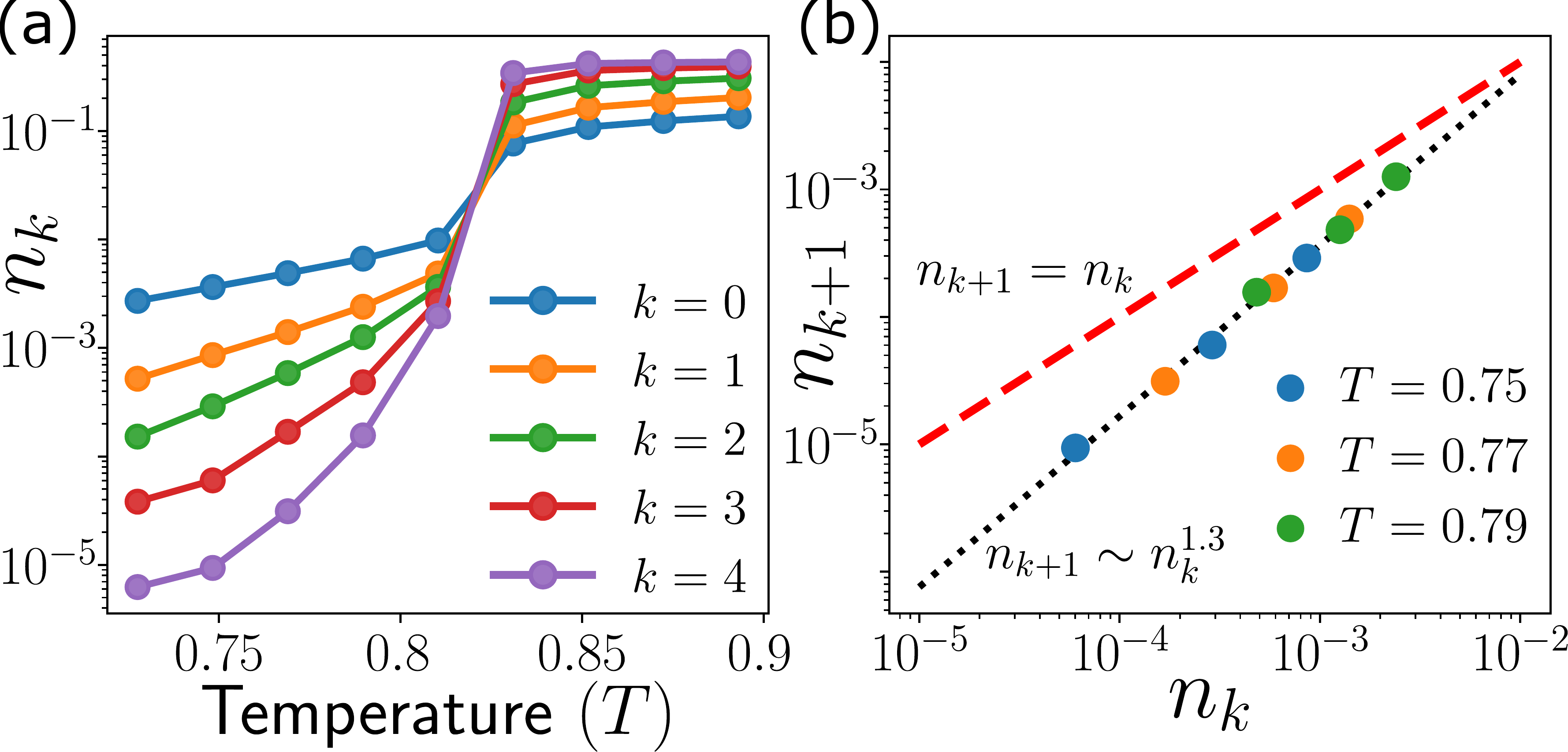}%
}
\caption{Quasi-local RG decoder for the Gibbs state of $H_\text{sq}$ with $\alpha = 2.5$ (a) Anyon density after $k$ iterations of the decoder for different temperatures. (b) In the low-temperature phase, the anyon density goes to zero as $n_{k+1} \sim (n_k)^\eta$, with $\eta \approx 1.3$ (fit and reference $n_{k+1} = n_{k}$ shown in black and red, respectively).} 
\label{fig:rgdecoder}
\end{figure}

The success of the RG decoder can be checked numerically by tracking the anyon density $n_k$ at each step. If $n_k$ decreases, the decoder approaches a toric code ground state. The rate at which $n_k$ decays is a measure of how quickly the defects are cleaned up. Following Ref.~\onlinecite{Sang2024}, when $n_{k+1} \sim (n_k)^\eta$ with $\eta>1$, the decoder can achieve arbitrary accuracy at quasi-local depth. We therefore use the flow of anyon density under the RG as a proxy measure for LRE of the Gibbs state.

The results of this numerical test are shown in Fig.~\ref{fig:rgdecoder} for $\alpha = 2.5$. For each temperature, we simulate $10^4$ equilibrated configurations of $\rho_\beta \propto e^{-\beta H_\text{sq}}$ in a system of size $L=128$. As before, we restrict our analysis to the magnetic sector and employ classical Monte Carlo.
We then apply four iterations of the RG decoder to each configuration, tracking the anyon density after each step. Unfortunately, the initial system size limits the number of RG steps that we can apply, but a clear trend is visible.

At low temperatures, the anyon density decreases, and the state tends to a toric code ground state. At high temperatures, the decoder increases the anyon density and therefore fails. The threshold of the decoder, which appears to be around $T = 0.82$ in Fig.~\ref{fig:rgdecoder}(a), provides a lower bound on the crossover temperature of the quantum memory. Note that the decoder threshold is quite close to the estimated $T_c$ from the specific heat (Fig.~\ref{fig:heat}) and the memory time (Fig.~\ref{fig:memory}). This is in stark contrast to the dephased toric code, for which known RG decoders fail well before the quantum memory is actually compromised (cf. Table 1 in Ref.~\onlinecite{Selub2026}). 
{The RG decoder is surprisingly successful in our long-range models because} anyons are both rare and tightly bound at low temperatures. 

Not only does the decoder successfully clean up anyons at low temperatures, it also does so quickly enough to be a quasi-local channel. In Fig.~\ref{fig:rgdecoder}(b) we show the anyon density at the $k$-th and $(k+1)$-th steps for three temperatures below the critical point. The density appears to decay as $n_{k+1} \sim (n_k)^\eta$ with $\eta \approx 1.3$. As a result, only range $\mathcal{O}(\ln L)$ operations are needed to map $\rho_\beta$ arbitrarily close to the logical subspace, providing {further} evidence that $\rho_\beta$ is LRE.

If the RG decoder were reversible, we could also conclude that $\rho_\beta$ belongs to the same mixed-state phase as the zero-temperature toric code~\cite{Sang2025}. {This is a stronger statement than LRE, as it requires a quasi-local channel that reconstructs $\rho_\beta$ from $\rho_\infty$.} 
The authors of Ref.~\onlinecite{Sang2024} argued that $C_\text{RG}$ is reversible when acting on the Gibbs state of the \emph{ordinary} toric code.
However, their argument relies on the assumption of uncorrelated anyons, which is not true in our long-range models.
We therefore leave the question of the mixed-state phase of $\rho_\beta$ to future work.

\section{Discussion}\label{sec:discussion}

The long-range toric code is, to the best of our knowledge,
the first example of a thermodynamically well-defined system in two dimensions that hosts finite-temperature topological order. The key ingredients in our models are large-loop stabilizer interactions that confine anyons while preserving the topologically ordered ground states of the toric code. We have characterized the finite-temperature topological order both dynamically through its quantum memory and in equilibrium, via the long-range entanglement of Gibbs states. In addition, we have discussed regimes that exhibit signatures of topological order only up to a large but finite system size, yet can still serve as effective quantum memories.

The energetics and dynamics of anyons in the $1 < \alpha \leq 2$ regime exhibit a striking parallel with those of defects  in 2D crystals~\cite{Pretko2018May}. Separating two anyons by an $\mathcal O(L)$ distance costs an energy $\Delta E\sim L^{3-\alpha}$, while the energy of a tightly bound anyon dipole diverges more slowly, as $L^{2-\alpha}$ for $1<\alpha<2$ and as $\ln L$ for $\alpha=2$. In contrast, compact anyon quadrupoles have finite excitation energies. This hierarchy of energy scales parallels that of disclinations, dislocations (disclination dipoles), and point defects such as vacancies and interstitials (disclination quadrupoles) in a 2D crystal, whose energies scale as $L^2$, $\ln L$, and $\mathcal O(1)$, respectively. 

The parallel between anyons in our models and crystalline defects
extends to their constrained dynamics~\cite{Pretko2020Mar,Nandkishore2019Mar}: individual anyons behave as immobile fractons, anyon dipoles as lineons with 1D mobility, and quadrupoles as fully mobile excitations. These mobility constraints respectively mirror those of disclinations, dislocations, and vacancies or interstitials in a crystal~\cite{Pretko2018May,pretko2019crystal,kim2024dynamical}.
This correspondence is reminiscent of the fracton-elasticity duality~\cite{Pretko2018May} and suggests a deeper connection between our long-range gauge theory and a modified form of elasticity. In particular, it raises the intriguing possibility that the finite-temperature transition out of the quantum memory phase may be understood as a defect-mediated melting transition. Exploring this possibility is an interesting direction for future research.

Our work opens several directions for future research. On the experimental side, quantum imaginary-time evolution~\cite{motta2020determining,mcardle2019variational} could provide a way to prepare the topologically ordered Gibbs states of our long-range toric code models on programmable quantum devices. Complementary equilibrium diagnostics, such as topological entanglement negativity~\cite{Lu2020}, could further characterize the long-range entanglement of these Gibbs states. More broadly, our construction provides a framework for exploring topological order in long-range interacting quantum many-body systems. In particular, long-range generalizations of the string-net models~\cite{levin2005string} could stabilize finite-temperature non-Abelian topological order. Exploring these possibilities may provide new routes toward robust topological quantum memories beyond the toric code.

\section*{Acknowledgments}
We thank Eduardo Fradkin, Amanda Gatto Lamas, Jintae Kim, Matteo Ippoliti, Tom Iadecola, Zhen Bi, Giuseppe De Tomasi, Akash Vijay, Jong Yeon Lee, and Steven Girvin for helpful discussions.
J.G. is supported by the C-TEQ Postdoctoral Fellowship at the Pennsylvania State University and acknowledges the hospitality of MPIPKS Dresden, where part of this work was conducted.
K.S.K. is supported by the Anthony J. Leggett Postdoctoral Fellowship at the University of Illinois Urbana-Champaign. Part of the work by K.S.K. was
performed at the Aspen Center for Physics, which is supported by a grant from the Simons Foundation (1161654,
Troyer). 

\section*{Code Availibility}
The code used to generate all numerical data supporting the findings of this article is openly available~\cite{code_github}.

\appendix
\section{Other long-range loop models}\label{app:other_loops}

In this appendix, we discuss LRTC models that contain more general loop stabilizers, such as rectangles and self-avoiding polygons. 
As in the square-loop model, these interactions preserve the toric-code ground state manifold while modifying the energetics of anyon excitations. 

\subsection{Rectangular loops}

A simple extension of the square-loop model is obtained by including rectangular loops on the direct and dual lattices, with coupling strengths that decay algebraically with their perimeter. The loop-counting arguments of Secs.~\ref{sec:anyon_energetics} and \ref{sec:energetics_finiteT} carry over with only a minor modification: the sum over the side length of a square is replaced by two sums over the side lengths $R_x$ and $R_y$ of a rectangle.
Consequently, the fully confining, partially confining, and deconfined regimes 
are shifted to $\alpha \mapsto \alpha+1$, with otherwise qualitatively similar finite-temperature behavior. For practical implementations in quantum simulators, however, it may be simpler to introduce only a few small rectangular stabilizers (e.g., $2\times1$ and $1\times2$ rectangles), rather than loops of arbitrarily large size, with the goal of enhancing the quantum memory lifetime.

\subsection{Self-avoiding polygon loops}

Another generalization includes stabilizers supported on self-avoiding polygons (SAP) on the lattice and dual lattice, denoted by $\mathcal{P}$ and $\tilde{\mathcal{P}}$, respectively. The SAP model is defined by the Hamiltonian:
\begin{equation}
\label{eq:ham_wiggle}
H_\text{SAP} = -\sum_{\mathfrak{p} \in \mathcal{P}} \frac{J_B}{\mu^{\ell(\mathfrak{p})}\,\ell(\mathfrak{p})^{\alpha_B}} B_{\mathfrak{p}} - \sum_{\tilde{\mathfrak{p}} \in \tilde{\mathcal{P}}} \frac{J_A}{\mu^{\ell(\tilde{\mathfrak{p}})}\,\ell(\tilde{\mathfrak{p}})^{\alpha_A}} A_{\tilde{\mathfrak{p}}}.
\end{equation}
Here, $\ell(\mathfrak{p})$ is the perimeter of the loop $\mathfrak{p}\in \mathcal{P}$. As in Sec.~\ref{sec:LRTC}, we assume exact eletromagnetic duality and take $J_A=J_B=1$ and $\alpha_A=\alpha_B=\alpha$ for simplicity. Once again, the loop stabilizers leave the toric code ground state manifold intact, but modify the energies of excited states.

There are two main differences compared to $H_\text{sq}$: the sums are now over all SAPs, and each term contains an extra exponentially decaying factor $1/\mu^{\ell(\mathfrak{p})}$. 
This factor is necessary to compensate the exponential growth in the number of SAPs on the lattice. Indeed, at large perimeter $\ell$, the number of closed, self-avoiding walks asymptote to~\cite{Nienhuis1982,Jensen1999}:
\begin{equation}\label{eq:sap}
\mathcal{N}_\text{SAP}(\ell) \sim \frac{\mu^\ell}{\ell^{5/2}},
\end{equation}
where the critical exponent $5/2$ is universal for all 2D lattices, while the connective constant $\mu$ is lattice dependent, with $\mu \approx 2.638$ on the square lattice.
Summing all possible loops gives a bandwidth of $E_\text{bw} \sim \sum_\ell \mathcal{N}_\text{SAP}(\ell)/(\mu^\ell \ell^\alpha)$, which using Eq.~\eqref{eq:sap} is extensive for $\alpha > -3/2$.

\begin{figure}[t!]
\centering
\includegraphics[width=\linewidth]{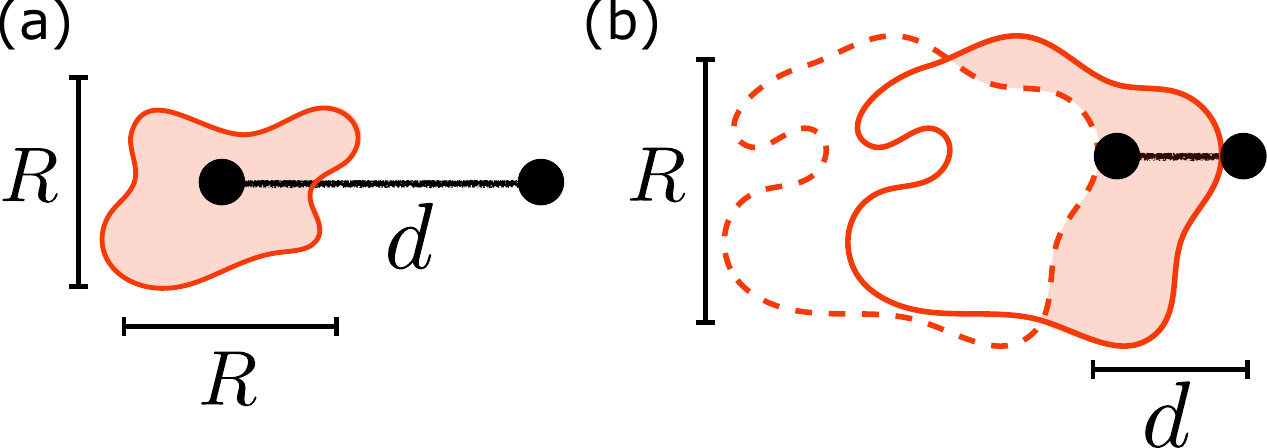}
\caption{Counting the number of SAPs of typical size $R$ that enclose a single anyon in a $d$-separated pair. (a) If $R < d$ the square can be translated $R$ units in both directions and still enclose the anyon. (b) If $R > d$ the square can be translated $R$ units vertically but only $d$ horizontally.
}
\label{fig:area_sap}
\end{figure}

We now calculate the anyon-pair energy in the SAP model using the loop-enclosure argument of Sec.~\ref{sec:anyon_energetics}. Only loops enclosing a single anyon contribute:
\begin{equation}
    \Delta E_\text{SAP}(\v{d})
    = \sum_{\mathfrak{p} \in \mathcal{P}} \frac{\delta_{n_\mathfrak{p},1}}{\mu^{\ell(\mathfrak{p})}\ell(\mathfrak{p})^\alpha}.
    \label{eq:deltaE_general_SAP}
\end{equation}
Here, $n_\mathfrak{p}$ is the number of anyons enclosed by the loop $\mathfrak{p}$. As in Sec.~\ref{sec:anyon_energetics}, we take two anyons separated by a horizontal vector $\v{d} = d \hat{\bm{x}}$. In this case, it is useful to decompose Eq.~\eqref{eq:deltaE_general_SAP} into separate sums over loop perimeter and loop shape:
\begin{equation}\label{eq:deltaE_wiggle}
    \Delta E_\text{SAP} ({d}) = \sum_{\ell}\frac{1}{\mu^\ell \ell^\alpha} \sum_{\mathfrak{p}_\ell \in \mathcal{P}_\ell} {N_\text{enc}(\mathfrak{p}_\ell, {d})}, 
\end{equation}
where $\mathcal{P}_\ell$ is the set of all self-avoiding loops of perimeter $\ell$. Unlike the square loop case, the number of loops that enclose one anyon also depends on the shape of the loop. Nevertheless, we can still interpret $N_\text{enc}(\mathfrak{p}_{\ell}, {d})$ as the area swept out by $\mathfrak{p}_\ell$ as it encloses one anyon, as depicted in Fig.~\ref{fig:area_sap}.

Once again, there are two regimes for this enclosing area. If the linear size of the loop $\mathfrak{p}_\ell$ is smaller than the anyon separation $d$, then $\mathfrak{p}_\ell$ can be translated across its full area while enclosing a single anyon (Fig.~\ref{fig:area_sap}(a)). On the other hand, if $\mathfrak{p}_\ell$ is larger than $d$, then the loop can only be translated a distance $d$ towards the external anyon before enclosing it as well (Fig.~\ref{fig:area_sap}(b)). 
In both regimes, $N_\text{enc}$ is related to the area of overlap between $\mathfrak{p}_\ell$ and a copy of $\mathfrak{p}_\ell$ translated by $d$. Denoting this intersection area by $A_\text{isct}(\mathfrak{p}_\ell, d)$ and the loop area by $A(\mathfrak{p}_\ell)$,
\begin{equation}
    \label{eq:area_shift}
    N_\text{enc}(\mathfrak{p}_\ell, d) = A(\mathfrak{p}_\ell) - A_\text{isct}(\mathfrak{p}_\ell, d).
\end{equation}
In the small-loop regime, $\mathfrak{p}_\ell$ and its translated partner do not intersect, and Eq.~\eqref{eq:area_shift} simply reduced to the area of the loop. In the large loop regime, only a fraction of the total area of $\mathfrak{p}_\ell$ contributes to $N_\text{enc}(\mathfrak{p}_\ell, d)$.

The energy calculation in Eq.~\eqref{eq:deltaE_wiggle} requires the area difference $N_\text{enc}$ to be averaged over all self-avoiding loops of perimeter $\ell$.
This is a daunting task, which we circumvent by appealing to a mean-field approximation. The typical area of a self-avoiding loop scales with its perimeter as $\ell^{3/2}$~\cite{Nienhuis1982}. Since $N_\text{enc}(\mathfrak{p}_\ell, d)$ is itself a difference of areas, it should exhibit a similar scaling. Consequently, we replace the average of $N_\text{enc}$ over all shapes
by a single $N_\text{enc}$ for the ``average shape''.
By average shape, we mean an isotropic loop of area $A_\text{typ}(\ell) \sim \ell^{3/2}$ and linear size $R_\text{typ}(\ell) \sim \ell^{3/4}$, the typical values for SAPs of length $\ell$. 

\begin{figure}[t!]
\centering
\makebox[\linewidth][c]{%
    \includegraphics[width=0.85\linewidth]{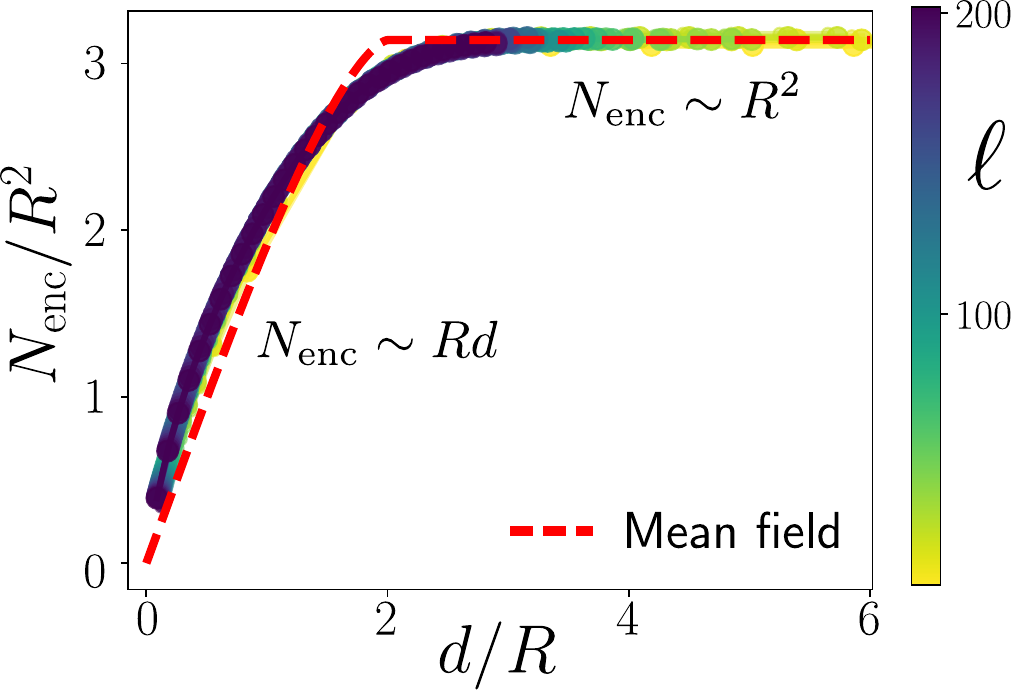}%
}
\caption{Monte Carlo sampling of the area $N_\text{enc}(\ell, d)$ for random self-avoiding polygons of perimeter $\ell$. The scaling of $N_\text{enc}$ is similar the mean-field estimate (red dashed line), which is based on a circle of radius $R = R_\text{typ}(\ell)$, the typical radius of gyration of a SAP of perimeter $\ell$. In the large loop regime (small $d/R$), $N_\text{enc}$ is slightly larger than the mean-field estimate, suggesting an even stronger anyon confinement than predicted by Eq.~\eqref{eq:deltaE_sap_scaling}.} 
\label{fig:meanfield}
\end{figure}

Using this approximation, the small and large loop regimes become very similar to the analogous regimes in the square loop model:
\begin{equation}\label{eq:Nenc_sap}
    {N_\text{enc}}(\ell, d) \sim 
    \begin{cases}
    R_\text{typ}(\ell)^2, & R_\text{typ}(\ell) < d, \\
    R_\text{typ}(\ell) d, & R_\text{typ}(\ell) > d. \\
    \end{cases}
\end{equation}
This expression has a simple geometric interpretation, shown in Fig.~\ref{fig:area_sap}: small loops can be translated $R_\text{typ}$ in both directions, while large loops can only be translated up to $d$ units towards the second anyon without also enclosing it. To confirm our mean-field picture, we 
use the pivot algorithm~\cite{Madras1990} to sample the overlap area in Eq.~\eqref{eq:area_shift} over random SAPs of fixed perimeter. The sampled value of $N_\text{enc}(\ell, d)$ is in good agreement with the value of $N_\text{enc}(\ell, d)$ for a circle of radius $R_\text{typ}(\ell)$, as shown in Fig.~\ref{fig:meanfield}, suggesting that Eq.~\eqref{eq:Nenc_sap} approximately captures the scaling of $N_\text{enc}$.

Returning to Eq.~\eqref{eq:deltaE_wiggle}, the mean field approximation yields
\begin{equation}\label{eq:deltaE_wiggle_mf}
    \Delta E_\text{SAP}(d) = \sum_\ell \frac{\mathcal{N}_\text{SAP}(\ell) N_\text{enc} (\ell, d)}{\mu^\ell \ell^\alpha},
\end{equation}
where $\mathcal{N}_\text{SAP}$ is the total number of self-avoiding loops of perimeter $\ell$. Inserting the scaling forms from Eqs.~\eqref{eq:sap} and \eqref{eq:Nenc_sap} and taking the continuum limit, we obtain
\begin{equation}\label{eq:deltaE_wiggle_int}
\begin{aligned}
    \Delta E_\text{SAP}(d) &
    \!\sim \!\int_1^{R_\text{typ}(\ell) = d} \frac{R_\text{typ}(\ell)^2}{\ell^{\alpha+5/2}}d\ell  + \!\! \int_{R_\text{typ}(\ell) = d}^{R_\text{typ}(\ell) = L}
    \! \frac{R_\text{typ}(\ell) d}{\ell^{\alpha + 5/2}} d\ell \\
    & \sim \int_1^{d^{4/3}} \frac{d\ell}{\ell^{\alpha + 1}} +  d \int_{d^{4/3}}^{L^{4/3}} \frac{d\ell}{\ell^{\alpha + 7/4}}.
\end{aligned}
\end{equation}
Although the Hamiltonian $H_\text{SAP}$ is extensive for $\alpha > -3/2$, we must further impose $\alpha > -3/4$ to obtain an anyon separation energy that is independent of $L$. As for the square loop model, there are three regimes:
\begin{equation}
    \label{eq:deltaE_sap_scaling}
    \Delta E_\text{SAP}(d) \sim
    \begin{cases}
        d^{-4\alpha/3}, & -3/4 < \alpha < 0 , \\
        \ln d, & \alpha = 0, \\
        \text{const}. & \alpha > 0.
    \end{cases}
\end{equation}
The fact that anyons are only confined when $\alpha < 0$, where that the power-law ``decay'' in the Hamiltonian is actually a power-law growth, may appear to be a problem. However, we must remember that terms in $H_\text{SAP}$ are also exponentially suppressed by loop perimeter. {We  note that Eq.~\eqref{eq:deltaE_sap_scaling} applies to isolated anyons: the finite-temperature effects discussed in Sec.~\ref{sec:energetics_finiteT} of the main text also alter the energetics of the SAP model.}

The case of $\alpha=0$, which is logarithmically confining, is especially interesting. Here loops are exponentially suppressed as $\mu^{-\ell}$ without an additional power-law. 
Such terms are generated perturbatively by the ordinary toric code in an external field
\begin{equation}\label{eq:ham_tc_field}
    H = -J_A \sum_v A_v - J_B \sum_p B_p - h \sum_i Z_i.
\end{equation}
The field $Z_i$ creates violations of the $A_v$ terms next to the link $i$. If $h \ll J_A$, perturbation theory in the low-energy subspace produces closed strings of $Z_\ell$ to preserve $A_v = +1$. At order $\ell$ in perturbation theory, these closed strings include precisely the self-avoiding loops of length $\ell$ that appear in $H_\text{SAP}$. Such terms come with a coefficient that scales as $(h/J_A)^\ell$, which for an appropriate choice of $h$ could be tuned to match $\mu^{-\ell}$. However, we have ignored combinatorial factors and other perturbatively-generated terms. Moreover, $\mu \approx 2.6$ for the square lattice, which would require $h\approx 0.4J_A$ --- well outside the perturbative regime.

\section{Scaling of contractible Wilson loops}\label{app:wilson}

Unlike noncontractible Wilson loops, which require 
a specific choice of global spatial topology, {\it contractible} Wilson loops detect more local correlations 
and are insensitive to the global topological sector.
In this appendix, we discuss contractible Wilson loops in the square-loop LRTC model and show that their thermal expectation values at low temperatures drastically differ from the area-law scaling of the ordinary toric code. 

We focus on the scaling of the Wilson loop
\begin{equation}\label{eq:wilson}
W_R = \prod_{i \in \partial \mathfrak s(R)} Z_i,
\end{equation}
where $\mathfrak s(R)$ is a square of side length $R$ in the regime $1 \ll R \ll L$.
At zero temperature, the Wilson loop distinguishes the topologically ordered phase of pure $\mathbb{Z}_2$ gauge theory, where $\expval{W_R} \sim e^{-\kappa R}$ exhibits a perimeter law, from the trivial phase, where $\expval{W_R} \sim e^{-\sigma R^2}$ exhibits an area law~\cite{Kogut1979Oct}..\footnote{A perimeter law does not necessarily imply topological order in gauge theories coupled to dynamical matter~\cite{Hastings2005Jul,Fradkin}.} 
At finite temperature, the thermal expectation value of $W_R$ is given by
\begin{equation}\label{eq:wilson_thermal}
    \expval{W_R}_\beta = \frac{\Tr[ e^{-\beta H} W_R]}{\Tr[e^{-\beta H}]}.
\end{equation}
In the ordinary toric code, $\expval{W_R}_\beta = (\tanh \beta)^{R^2}$ in the thermodynamic limit $L \to \infty $, which decays exponentially with the area of the Wilson loop at any finite temperature~\cite{Wegner1971, Kogut1979Oct}. This is the same scaling as in the trivial phase of $\mathbb{Z}_2$ gauge theory, highlighting the loss of topological order at $T>0$. 

In contrast, three regimes of the square-loop LRTC 
show distinct behaviors at different length scales: 
(1) For $\alpha >3$, the Wilson loop scaling crosses over from  area-law behavior at $1 \ll R \ll L_{\rm cross}$ to 
power-law behavior at $L_{\rm cross} \ll R \ll L$.
Here, $L_{\rm cross} \sim e^{\Delta/2T}$ is the crossover length scale above which anyons proliferate, where $\Delta$ is the energy cost of an isolated anyon.
(2) For $2  < \alpha \leq 3$, the Wilson loop scaling crosses over from perimeter-law behavior at $1 \ll R \ll L_{\rm conf}$ to power-law behavior at $L_{\rm conf} \ll R \ll L$. 
(3) For $1 < \alpha \leq 2$, we argue based on the dilute quadrupole gas picture that the thermal expectation value of the Wilson loop approaches constant for $1 \ll R \ll L$. This constant Wilson loop conjecture is also corroborated by a rigorous upper bound. These scaling behaviors are summarized in Table \ref{tab:wilson_scaling}. 

\begin{table}[t]
\centering
\renewcommand{\arraystretch}{1.5}
\begin{tabular*}{\columnwidth}{@{\extracolsep{\fill}}lcc@{}}
\hline\hline
 & Intermediate scales & Large-$R$ asymptotics \\
\hline
$1<\alpha\leq2$
& \multicolumn{2}{c}{$\mathrm{Const.}>0\ ^*$} \\

$2<\alpha\leq3$
& Perimeter law
& Power law $\sim \beta/ R^{\alpha}$ \\

$\alpha>3$
& Area law
& Power law $\sim \beta/ R^{\alpha}$ \\
\hline\hline
\end{tabular*}
\caption{
Low-temperature scaling of contractible square Wilson loops of side length $R$ in the square-loop LRTC for $1\ll R\ll L$. The intermediate-scale perimeter and area laws cross over to asymptotic power-law decay for $\alpha>2$. $^*$The conjectured positive constant at sufficiently low $T$ is motivated by the dilute quadrupole gas picture and a rigorous upper bound.
}
\label{tab:wilson_scaling}
\end{table}

\subsection{Intermediate-scale area and perimeter laws for $\alpha > 2$}  

For $2<\alpha\leq3$, a dilute gas of tightly bound anyon dipoles provides a useful approximation at low temperatures. As discussed in Sec.~\ref{sec:energetics_finiteT}, a dipole changes the sign of a Wilson loop only when it crosses the loop boundary. Consequently, the number of dipoles contributing to the Wilson loop scales with its perimeter, giving rise to the approximate perimeter law $\langle W_R\rangle_\beta\sim e^{-4R n_{\rm pair}}$. A more careful treatment shows that $\langle W_R\rangle_\beta\sim e^{-4R\kappa(T)}$, with $\kappa(T)\sim n_{\rm pair}\xi$, where $\xi$ is the typical dipole size. The extra factor of $\xi$ comes from the number of ways a dipole of size $\xi$ straddles a given point of the loop boundary.
As we will see in the next subsection, this behavior holds at intermediate length scales $1\ll R\ll L_{\rm conf}$. At sufficiently large $R$, $L_{\rm conf} \ll R \ll L$, the perimeter law is replaced by a power-law decay, which can be established using rigorous lower and upper bounds.

For $\alpha>3$, on the other hand, anyons are deconfined and can be approximated at low temperatures as a dilute gas of uncorrelated individual anyons. Since the Wilson loop measures the parity of the anyons it encloses, its expectation value decays exponentially with the enclosed area, $\langle W_R\rangle_\beta\sim e^{-\sigma(T)R^2}$, where $\sigma(T)\approx 2n_{\rm a}$, with $n_{\rm a}$ the anyon density. 
Again, at sufficiently large length scales $1/\sqrt{\sigma} \ll R \ll L $, the area-law contribution becomes exponentially small, and the Wilson loop eventually crosses over to the asymptotic power-law decay established in the next subsection.

\subsection{Asymptotic power-law decay for $\alpha > 2$}

The perimeter- and area-law Wilson loop behaviors discussed above are a transient behavior at intermediate length scales. We now establish that for $\alpha >2$, the Wilson loop expectation value in our LRTC model exhibits a much slower  algebraic decay at asymptotically large $R$ at any finite temperature
\begin{align}
  \expval{W_R}_\beta \sim 
  \frac{\beta}{(4R)^\alpha},
\end{align}
up to a constant prefactor. 
We establish this result by obtaining lower and upper bounds with matching asymptotic scaling. Interestingly, this algebraic decay evades a perimeter-law bound derived for general lattice gauge theories~\cite{Simon1982} by relaxing the underlying locality assumption.

Both lower and upper bounds can be established by using  the generalized Griffiths-Kelly-Sherman (GKS) inequalities~\cite{griffiths1967correlations,kelly1968general}, which apply to classical Ising models with ferromagnetic multispin interactions.
Consider the Hamiltonian:
\begin{align}
\label{eq:general_classical_Ising}
    H = - \sum_{A} J_A \Sigma_A, \ \ \ 
    \Sigma_A = \prod_{i \in A} \sigma _i, \ \ \ J_A \geq 0,
\end{align}
where $\sigma_i = \pm 1 $ and  $A$ denotes an arbitrary subset of spins, including a single spin.
The GKS inequalities state that thermal expectation values satisfy
\begin{align}
    \expval{\Sigma_A} \geq 0, 
\end{align}
\begin{align}
    \expval{\Sigma_A \Sigma_B} 
    -\expval{\Sigma_A}  \expval{\Sigma_B} \geq 0.
\end{align}
The latter, in particular, implies
\begin{align}
\label{eq:GKS_monotonicity}
    \frac{\partial \expval{\Sigma _A} }{\partial J_B} = \beta 
    \left [ \expval{\Sigma_A \Sigma_B} 
    -\expval{\Sigma_A}  \expval{\Sigma_B} 
    \right ]
    \geq 0.
\end{align}
In other words, increasing any positive coupling cannot decrease single- or multi-spin expectation values: we will refer to this as Griffiths monotonicity in the following.

\smallskip
Since \(H_{\rm sq}\) is a CSS Hamiltonian, the expectation value of \(W_R\) is determined entirely by its classical magnetic ($Z$) sector, to which the GKS inequalities apply. 
The lower bound for $\expval{W_R}$ is then a direct consequence of the Griffiths monotonicity and the fact that the LRTC Hamiltonian contains a term proportional to  $W_R.$ By removing all other terms in the magnetic sector, we obtain
\begin{align}
    \expval{W_R} &\geq  \expval{W_R}
    \vert_{H= - W_R/(4R)^\alpha}
    \nonumber 
    \\
    & = \tanh[\beta/(4R)^\alpha] \sim 
    \beta/(4R)^\alpha.
\end{align}

The upper bound of the Wilson loop expectation value is given by Ising correlation functions in the one-dimensional long-range Ising model, as we show in the following. For simplicity, we consider open boundary conditions and take the thermodynamic limit before taking $R\to\infty$. The analysis extends to periodic boundary conditions with an additional treatment of the global even-anyon-parity constraint. First, we choose a horizontal row \(\mathcal R\) containing one side of $W_R$ and apply a positive Zeeman field $h$ to all spins outside this row:
\begin{align}
    H_{\rm row } (h)  =
    -\sum_{\mathfrak s \in \mathcal S } 
    \frac 1 {\ell(\mathfrak s)^\alpha }\mathcal B_{\mathfrak s} 
    - h \sum_{i\notin \mathcal R } Z_i.
\end{align}
Griffiths monotonicity implies 
$\expval{W_R}  \leq \expval{W_R}_{h},$
where the right-hand side is the thermal expectation value with respect to $H_{\rm row}(h)$. Taking $h\to \infty$, we obtain
\begin{align}
    \expval{W_R} \leq \expval{W_R}_{h = \infty}.
\end{align}
In this limit, all spins outside of row $\mathcal R$ are frozen to $Z_{i} = 1$, and the Hamiltonian reduces to a one-dimensional problem:
\begin{align}
    H_{\rm row } (h=\infty)  =
    - 2 \sum _x\sum_{R= 1}^\infty 
    \frac 1 {(4 R)^\alpha} \prod_{i=x}^{x+ R - 1} Z_i ,
\end{align}
up to an additive constant (a factor of two comes from contributions from two square loops, one above and one below the row).
We now introduce new Ising spins $\sigma_i $ via $Z_i = \sigma_i \sigma_{i+1}$, where \(\sigma_i\) are defined on the vertices along the row, with one boundary spin fixed to \(\sigma_0=1\). Then, 
\begin{align}
    H_{\rm row } (h=\infty)  =
    - 2 \sum _x\sum_{R= 1}^\infty 
    \frac {\sigma_x \sigma_{x + R }} {(4 R)^\alpha}  .
\end{align}
This is the Hamiltonian of a one-dimensional long-range Ising model. Under the same mapping, $W_R= \sigma_x \sigma_{x+R}$, so
\begin{align}
    \langle W_R\rangle
    \leq\langle\sigma_x\sigma_{x+R}\rangle_{\rm LRI}.
\end{align}
For $\alpha >2$, the long-range Ising model is disordered at any finite temperatures, and its two-point correlation function decays algebraically at asymptotically large distance~\cite{spohn1999decay,aoun2021sharp}:
\begin{align}
    \expval{\sigma_x \sigma_{x+R}}_{\rm LRI}  \sim \frac{2 \beta \chi_\beta^2 }{(4 R)^\alpha},
\end{align}
where $\chi_\beta = \sum_{y} \expval {\sigma_x \sigma_y }_{\rm LRI}$ is the dimensionless zero-field susceptibility, which is finite.
Therefore, we have established the power-law upper bound 
\begin{align}
    \expval{W_R}_{\rm LRI} \leq 
    \expval{\sigma_x \sigma_{x+R}}_\beta  \sim \frac{2 \beta \chi_\beta^2 }{(4 R)^\alpha}.
\end{align}
Since the lower and upper bounds exhibit the same asymptotic scaling, we conclude that the Wilson loop exhibits an asymptotic power-law decay for $\alpha > 2$ and $T>0$.

As discussed in Sec.~\ref{sec:energetics_finiteT}, this asymptotic power-law decay implies that the anyon separation energy approaches a finite limit as $d \to \infty$, $\Delta E(d)\sim\Delta E^\infty-d^{3-2\alpha}$, confirming that confinement is screened at large distances for $\alpha >2 $.

\subsection{Conjectured  Wilson loop plateau for $1 < \alpha \leq 2$}

For $1<\alpha\leq2$, the fundamental finite-energy excitations are tightly bound anyon quadrupoles. As discussed in Sec.~\ref{sec:energetics_finiteT}, a quadrupole crossing a straight segment of a square Wilson loop does not change its sign, since it addes an even number of anyons inside the loop. Only quadrupoles that straddle one of the four corners can flip the Wilson loop. Approximating the low-temperature background by a dilute gas of independent elementary quadrupoles with density $n_{\rm quad}$, we therefore obtain 
$\langle W_R\rangle_\beta\approx(1-2n_{\rm quad})^4\approx e^{-8n_{\rm quad}}$, as in Eq.~\eqref{eq:wilson_constant}. Since this expression is independent of $R$, it suggests that the Wilson loop expectation value approaches a nonzero constant at sufficiently low temperatures. At a critical temperature where anyon dipoles proliferate, this constant plateau would be replaced by the asymptotic power-law scaling discussed in the previous subsection.

\begin{figure*}[t!]
\centering
\includegraphics[width=0.93\textwidth]{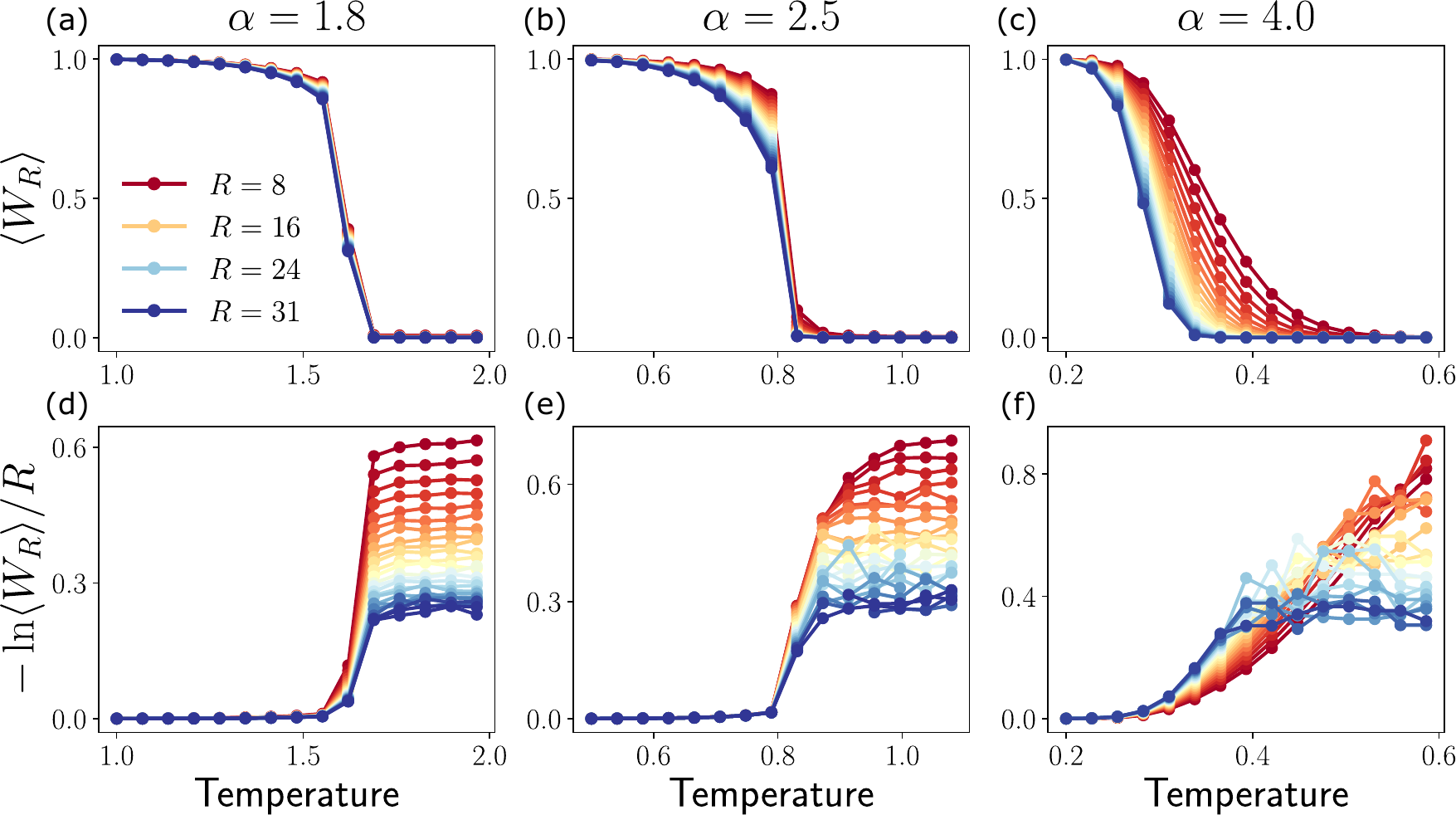}
\caption{Contractible Wilson loop expectation values. All quantities are sampled using classical Monte Carlo simulations of system of size $L=64$, and square Wilson loops of side length $8\leq R \leq 31$. (a, d) Wilson loops for $\alpha = 1.8$ are approximately $R$-independent at low temperatures, while they decay with $R$ more slowly than a perimeter law at high temperatures. (b, e) For $\alpha = 2.5$, an approximate perimeter law at low temperatures gives way to a slower-than-perimeter law behavior at higher temperatures. (c, f) For $\alpha = 4$, an approximate area law at low temperatures gives way to a slower-than-perimeter law behavior at higher temperatures.}
\label{fig:wilson}
\end{figure*}

We can also establish a rigorous upper bound that is independent of $R$, using the same analysis presented in the previous subsection. For $1<\alpha\leq2$, the one-dimensional long-range Ising model exhibits a ferromagnetically ordered phase below a critical temperature $T_c^{\rm LRI}$~\cite{Dyson1969,frohlich1982phase}. In this phase, the two-point correlation function approaches a nonzero constant at large distances $R \to \infty$, giving
\begin{align}
     \langle W_R\rangle
    \leq
    \langle\sigma_x\sigma_{x+R}\rangle_{\rm LRI}
    \sim m_T^2 ,
\end{align}
where $m_T\neq 0$ is the spontaneous magnetization of the long-range Ising model below $T_c^{\rm LRI}$. This constant upper bound is consistent with the conjectured Wilson loop plateau. However, we have not been able to establish a matching nonzero lower bound, and leave a rigorous proof of the plateau to future work.

\subsection{Numerical Results}

We now corroborate our analytical predictions of Wilson loop scaling with Monte Carlo simulations.
Following a similar procedure as in the main text, we generate $10^4$ samples of the thermal state of $H_\text{sq}$ for a range of temperatures and exponents $\alpha$. In Fig.~\ref{fig:wilson}, we show the Wilson loop expectation value $\langle W_R\rangle$ (a-c) and the quantity $-\ln\langle W_R\rangle/R$ (d-e), for different contractible loop sizes $R$, for $\alpha=1.8,\ 2.5$, and $4.0$. The latter quantity is independent of $R$ for a perimeter law, increases with $R$ for an area law, and decreases for a power law.

For $\alpha=2.5$, the low-temperature behavior ($T\lesssim 0.8$) is consistent with an intermediate-scale perimeter law, while at higher temperatures, the Wilson loop exhibits a slower-than-perimeter-law decay, consistent with the asymptotic power law derived above. This crossover can be understood from the temperature dependence of the confinement length $L_{\rm conf}\sim e^{\beta\Delta_2}$, which exponentially decreases with increasing temperature. At low temperatures, the accessible loop sizes satisfy $R\ll L_{\rm conf}$, whereas at higher temperatures, the asymptotic power-law regime becomes accessible within our finite-size simulations.

For $\alpha=4$, the Wilson loop instead exhibits an approximate area law at low temperatures, consistent with the presence of deconfined anyons. As shown in Fig.~\ref{fig:wilson}(f), $-\ln\langle W_R\rangle/R$ initially increases with $R$, as expected for an area law. At higher temperatures, however, this quantity decreases with $R$, indicating a slower-than-perimeter-law decay consistent with the asymptotic power law. Again,  as the temperature increases, the density of thermally excited anyons grows, suppressing the area-law contribution and allowing the power-law tail to become visible at accessible loop sizes.

Finally, for $\alpha=1.8$, the Wilson-loop expectation value remains close to unity and shows little dependence on $R$ at low temperatures, consistent with the conjectured constant Wilson-loop plateau. Upon increasing temperature, $\langle W_R\rangle$ drops sharply, and $-\ln\langle W_R\rangle/R$ decreases with increasing $R$, suggesting a transition to a slower-than-perimeter-law decay consistent with power-law behavior. These results support the dilute-quadrupole picture of the low-temperature phase, although the limited system sizes do not allow us to establish the asymptotic scaling conclusively.

\section{Details of Monte Carlo simulations}\label{app:montecarlo}

In the main text, we perform several classical Monte Carlo simulations of the magnetic sector of $H_\text{sq}$. We briefly outline the procedures used below. There are two classes of simulations: cooling from an infinite temperature state and heating from a ground state. The former samples the full Gibbs distribution $\rho \propto e^{-\beta H_\text{sq}}$ at temperature $T = 1/\beta$ and maximally mixes all four logical sectors. The latter samples the Gibbs state within a single logical sector~\cite{Placke2024ldpc, Bergamaschi2025}, provided the temperature is low enough that this state remains stable.

The simulations are performed on an $L \times L$ square lattice with periodic boundary conditions in both directions. The coefficients in Eq.~\eqref{eq:ham_square} are taken to be $J_{A,B} = 4^\alpha \zeta(2)$, where $\zeta(2)$ is the Riemann zeta function (with this choice, the energy cost of two nearest-neighbor anyons is the same in $H_\text{sq}(\alpha = 3)$ and $H_\text{tc}$, so we have comparable energy scales to the ordinary toric code). The links of the lattice contain classical spins $s_i = \pm 1$. For cooling, the spins are initialized randomly to start in a $T=\infty$ state. For heating from a ground state, the spins are all initialized in the $+1$ state, representing a ground state in the logical sector with no noncontractible loops. 

At each update step, a single link is chosen and provisionally flipped. The energy difference $\Delta E$ between the pre- and post-flip configurations is calculated and the flip is accepted with a Metropolis probability $p = \text{min}(1, e^{-\beta \Delta E})$. To calculate this energy, we track all possible square stabilizers (of side length $R \in [1, L/2]$) that contain the link, and thus change sign when it is flipped. One time step corresponds to $2L^2$ attempted flips, i.e., once every spin has had a chance to update. For $1 < \alpha \leq 2$, we also include more general local updates, such as flipping two adjacent links, which corresponds to the creation of an anyon quadrupole. This is to ensure that the system properly thermalizes in this regime: a single anyon dipole cannot be injected as it has an energy diverging with the system size whereas anyon quadrupoles have a finite energy.

Thermodynamic expectation values like the specific heat are sampled by cooling from infinite temperature. In all cases, local equilibration and autocorrelation times are on the order of 10 Monte Carlo steps. For any measured quantity, $10^4$ equilibrated samples of a system are generated at each temperature. Quantities related to the memory stored in the ground state subspace, on the other hand, are sampled by heating a ground state. The probability of logical failure is sampled by decoding configurations after $200$ Monte Carlo steps, while the memory lifetime $t_\text{mem}$ is the average first time that the decoder fails to return the original ground state. The MWPM decoder used to define memory quantities is implemented using the PyMatching package~\cite{Higgott2025}.

\AddToHook{env/thebibliography/begin}{\raggedright}
\bibliography{ref.bib}

@misc{code_github,
  title = {{Code for numerical simulations of the long-range toric code}},
  author = {},
  year = {2026},
  howpublished = {\href{https://github.com/jgliozzi/LongRangeToricCode}{github.com/jgliozzi/LongRangeToricCode}}
  }

@article{Kitaev2003Jan,
  author = {Kitaev, A. {\relax Yu}.},
  title = {{Fault-tolerant quantum computation by anyons}},
  journal = {Ann. Phys.},
  volume = {303},
  number = {1},
  pages = {2--30},
  year = {2003},
  month = jan,
  issn = {0003-4916},
  publisher = {Academic Press},
  doi = {10.1016/S0003-4916(02)00018-0}
}

@article{Pretko2018May,
	author = {Pretko, Michael and Radzihovsky, Leo},
	title = {{Fracton-Elasticity Duality}},
	journal = {Phys. Rev. Lett.},
	volume = {120},
	number = {19},
	pages = {195301},
	year = {2018},
	month = may,
	publisher = {American Physical Society},
	doi = {10.1103/PhysRevLett.120.195301}
}

@article{pretko2019crystal,
  title={Crystal-to-fracton tensor gauge theory dualities},
  author={Pretko, Michael and Zhai, Zhengzheng and Radzihovsky, Leo},
  journal={Physical Review B},
  volume={100},
  number={13},
  pages={134113},
  year={2019},
  publisher={APS}
}

@article{Stahl2026Sep,
	author = {Stahl, Charles and Placke, Benedikt and Khemani, Vedika and Li, Yaodong},
	title = {{Slow mixing and emergent one-form symmetries in three-dimensional Z2 gauge theory}},
	journal = {Phys. Rev. E},
	volume = {114},
	number = {3},
	pages = {034126},
	year = {2026},
	month = sep,
	publisher = {American Physical Society},
	doi = {10.1103/5ypm-x96s}
}

@article{Nandkishore2019Mar,
	author = {Nandkishore, Rahul M. and Hermele, Michael},
	title = {{Fractons}},
	journal = {Annu. Rev. Condens. Matter Phys.},
	number = {Volume 10, 2019},
	pages = {295--313},
	year = {2019},
	month = mar,
	publisher = {Annual Reviews},
	doi = {10.1146/annurev-conmatphys-031218-013604}
}

@article{Pretko2020Mar,
	author = {Pretko, Michael and Chen, Xie and You, Yizhi},
	title = {{Fracton phases of matter}},
	journal = {Int. J. Mod. Phys. A},
	volume = {35},
	number = {06},
	year = {2020},
	month = mar,
	publisher = {World Scientific Publishing Company},
	doi = {10.1142/S0217751X20300033}
}

@article{Selub2026,
	author = {Selub, Nathaniel and Bhardwaj, Aditya and Lake, Ethan},
	title = {{Local decoders for fault-tolerant quantum computation and translation-invariant stabilizer codes}},
	journal = {arXiv},
	year = {2026},
	month = sep,
	eprint = {2609.11457},
}

@article{Wang2025Jan,
	author = {Wang, Zijian and Wu, Zhengzhi and Wang, Zhong},
	title = {{Intrinsic Mixed-State Topological Order}},
	journal = {PRX Quantum},
	volume = {6},
	number = {1},
	pages = {010314},
	year = {2025},
	month = jan,
	publisher = {American Physical Society},
	doi = {10.1103/PRXQuantum.6.010314}
}

@article{Hastings2005Jul,
	author = {Hastings, M. B. and Wen, Xiao-Gang},
	title = {{Quasiadiabatic continuation of quantum states: The stability of topological ground-state degeneracy and emergent gauge invariance}},
	journal = {Phys. Rev. B},
	volume = {72},
	number = {4},
	pages = {045141},
	year = {2005},
	month = jul,
	publisher = {American Physical Society},
	doi = {10.1103/PhysRevB.72.045141}
}

@article{Wegner1971,
	author = {Wegner, Franz J.},
	title = {{Duality in Generalized Ising Models and Phase Transitions without Local Order Parameters}},
	journal = {J. Math. Phys.},
	volume = {12},
	number = {10},
	pages = {2259--2272},
	year = {1971},
	month = oct,
	issn = {0022-2488},
	publisher = {AIP Publishing},
	doi = {10.1063/1.1665530}
}

@article{Nussinov2008Feb,
  author = {Nussinov, Zohar and Ortiz, Gerardo},
  title = {{Autocorrelations and thermal fragility of anyonic loops in topologically quantum ordered systems}},
  journal = {Phys. Rev. B},
  volume = {77},
  number = {6},
  pages = {064302},
  year = {2008},
  month = feb,
  publisher = {American Physical Society},
  doi = {10.1103/PhysRevB.77.064302}
}

@article{Chen2010Oct,
	author = {Chen, Xie and Gu, Zheng-Cheng and Wen, Xiao-Gang},
	title = {{Local unitary transformation, long-range quantum entanglement, wave function renormalization, and topological order}},
	journal = {Phys. Rev. B},
	volume = {82},
	number = {15},
	pages = {155138},
	year = {2010},
	month = oct,
	publisher = {American Physical Society},
	doi = {10.1103/PhysRevB.82.155138}
}

@book{Fradkin,
  author    = {Eduardo Fradkin},
  title     = {Quantum Field Theory: An Integrated Approach},
  publisher = {Princeton University Press},
  year      = {2021},
  isbn      = {9780691149080}
}

@article{Calderbank1996,
	author = {Calderbank, A. R. and Shor, Peter W.},
	title = {{Good quantum error-correcting codes exist}},
	journal = {Phys. Rev. A},
	volume = {54},
	number = {2},
	pages = {1098},
	year = {1996},
	month = aug,
	publisher = {American Physical Society},
	doi = {10.1103/PhysRevA.54.1098}
}

@article{Steane1996,
	author = {Steane, Andrew},
	title = {{Multiple-particle interference and quantum error correction}},
	journal = {Proc. R. Soc. A},
	volume = {452},
	number = {1954},
	pages = {2551--2577},
	year = {1996},
	month = nov,
	issn = {1364-5021},
	publisher = {The Royal Society},
	doi = {10.1098/rspa.1996.0136}
}

@article{Kogut1979Oct,
	author = {Kogut, John B.},
	title = {{An introduction to lattice gauge theory and spin systems}},
	journal = {Rev. Mod. Phys.},
	volume = {51},
	number = {4},
	pages = {659},
	year = {1979},
	month = oct,
	publisher = {American Physical Society},
	doi = {10.1103/RevModPhys.51.659}
}

@article{Simon1982,
	author = {Simon, B. and Yaffe, L. G.},
	title = {{Rigorous perimeter law upper bound on Wilson loops}},
	journal = {Phys. Lett. B},
	volume = {115},
	number = {2},
	pages = {145--147},
	year = {1982},
	month = aug,
	issn = {0370-2693},
	publisher = {North-Holland},
	doi = {10.1016/0370-2693(82)90815-2}
}

@article{Placke2024ldpc,
  author = {Placke, Benedikt and Rakovszky, Tibor and Breuckmann, Nikolas P. and Khemani, Vedika},
  title = {{Topological Quantum Spin Glass Order and its realization in qLDPC codes}},
  journal = {arXiv},
  year = {2024},
  month = dec,
  eprint = {2412.13248},
}

@article{Hong2025ldpc,
  author = {Hong, Yifan and Guo, Jinkang and Lucas, Andrew},
  title = {{Quantum memory at nonzero temperature in a thermodynamically trivial system}},
  journal = {Nat. Commun.},
  volume = {16},
  number = {316},
  pages = {316},
  year = {2025},
  month = jan,
  issn = {2041-1723},
  publisher = {Nature Publishing Group},
  doi = {10.1038/s41467-024-55570-7}
}

@article{Jensen1999,
  author = {Jensen, Iwan and Guttmann, Anthony J.},
  title = {{Self-avoiding polygons on the square lattice}},
  journal = {J. Phys. A: Math. Gen.},
  volume = {32},
  number = {26},
  pages = {4867},
  year = {1999},
  month = jul,
  issn = {0305-4470},
  publisher = {IOP Publishing},
  doi = {10.1088/0305-4470/32/26/305}
}

@article{Aguado2008Feb,
	author = {Aguado, Miguel and Vidal, Guifr{\ifmmode\acute{e}\else\'{e}\fi}},
	title = {{Entanglement Renormalization and Topological Order}},
	journal = {Phys. Rev. Lett.},
	volume = {100},
	number = {7},
	pages = {070404},
	year = {2008},
	month = feb,
	publisher = {American Physical Society},
	doi = {10.1103/PhysRevLett.100.070404}
}

@article{Alicki2009,
  author = {Alicki, R. and Fannes, M. and Horodecki, M.},
  title = {{On thermalization in Kitaev's 2D model}},
  journal = {J. Phys. A: Math. Theor.},
  volume = {42},
  number = {6},
  pages = {065303},
  year = {2009},
  month = jan,
  issn = {1751-8121},
  publisher = {IOP Publishing},
  doi = {10.1088/1751-8113/42/6/065303}
}

@article{Alicki2012,
  author = {Alicki, R. and Horodecki, M. and Horodecki, P. and Horodecki, R.},
  title = {{On Thermal Stability of Topological Qubit in Kitaev's 4D Model}},
  journal = {Open Systems {\&} Information Dynamics},
  volume = {17},
  number = {01},
  year = {2012},
  month = jan,
  publisher = {World Scientific Publishing Company},
  doi = {10.1142/S1230161210000023}
}

@article{Balasubramanian2026,
  author = {Balasubramanian, Shankar and Davydova, Margarita and Lin, Ting-Chun},
  title = {{A passive self-correcting quantum memory in three dimensions}},
  journal = {arXiv},
  year = {2026},
  month = may,
  eprint = {2605.10943},
}

@article{Higgott2025,
  author = {Higgott, Oscar and Gidney, Craig},
  title = {{Sparse Blossom: correcting a million errors per core second with minimum-weight matching}},
  journal = {Quantum},
  volume = {9},
  pages = {1600},
  year = {2025},
  month = jan,
  publisher = {Verein zur F{\ifmmode\ddot{o}\else\"{o}\fi}rderung des Open Access Publizierens in den Quantenwissenschaften},
  eprint = {2303.15933v2},
  doi = {10.22331/q-2025-01-20-1600}
}

@article{Breuckmann2021,
  author = {Breuckmann, Nikolas P. and Eberhardt, Jens Niklas},
  title = {{Quantum Low-Density Parity-Check Codes}},
  journal = {PRX Quantum},
  volume = {2},
  number = {4},
  pages = {040101},
  year = {2021},
  month = oct,
  publisher = {American Physical Society},
  doi = {10.1103/PRXQuantum.2.040101}
}

@article{Bergamaschi2025,
  author = {Bergamaschi, Thiago and Gheissari, Reza and Liu, Yunchao},
  title = {{Rapid mixing for Gibbs states within a logical sector: a dynamical view of self-correcting quantum memories}},
  journal = {arXiv},
  year = {2025},
  month = jul,
  eprint = {2507.10976},
}

@article{Sang2024,
  author = {Sang, Shengqi and Zou, Yijian and Hsieh, Timothy H.},
  title = {{Mixed-State Quantum Phases: Renormalization and Quantum Error Correction}},
  journal = {Phys. Rev. X},
  volume = {14},
  number = {3},
  pages = {031044},
  year = {2024},
  month = sep,
  publisher = {American Physical Society},
  doi = {10.1103/PhysRevX.14.031044}
}

@article{Sang2025,
  author = {Sang, Shengqi and Lessa, Leonardo A. and Mong, Roger S. K. and Grover, Tarun and Wang, Chong and Hsieh, Timothy H.},
  title = {{Mixed-state phases from local reversibility}},
  journal = {arXiv},
  year = {2025},
  month = jul,
  eprint = {2507.02292},
}

@article{Li2025Jun,
	author = {Li, Zhi and Lee, Dongjin and Yoshida, Beni},
	title = {{How Much Entanglement Is Needed for Topological Codes and Mixed States with Anomalous Symmetry?}},
	journal = {Phys. Rev. X},
	volume = {15},
	number = {2},
	pages = {021090},
	year = {2025},
	month = jun,
	publisher = {American Physical Society},
	doi = {10.1103/pw12-kdjx}
}

@article{Lessa2025Mar,
	author = {Lessa, Leonardo A. and Sang, Shengqi and Lu, Tsung-Cheng and Hsieh, Timothy H. and Wang, Chong},
	title = {{Higher-form anomaly and long-range entanglement of mixed states}},
	journal = {arXiv},
	year = {2025},
	month = mar,
	eprint = {2503.12792},
}

@article{Bravyi2013haah,
  author = {Bravyi, Sergey and Haah, Jeongwan},
  title = {{Quantum Self-Correction in the 3D Cubic Code Model}},
  journal = {Phys. Rev. Lett.},
  volume = {111},
  number = {20},
  pages = {200501},
  year = {2013},
  month = nov,
  publisher = {American Physical Society},
  doi = {10.1103/PhysRevLett.111.200501}
}

@article{Dyson1969,
  author = {Dyson, Freeman J.},
  title = {{Existence of a phase-transition in a one-dimensional Ising ferromagnet}},
  journal = {Commun. Math. Phys.},
  volume = {12},
  number = {2},
  pages = {91--107},
  year = {1969},
  month = jun,
  issn = {1432-0916},
  publisher = {Springer-Verlag},
  doi = {10.1007/BF01645907}
}

@article{thouless1969long,
  title={{Long-range order in one-dimensional Ising systems}},
  author={Thouless, DJ},
  journal={Physical Review},
  volume={187},
  number={2},
  pages={732},
  year={1969},
  publisher={APS}
}

@article{frohlich1982phase,
  title   = {{The Phase Transition in the One-Dimensional Ising Model with $1/r^2$ Interaction Energy}},
  author  = {Fr{\"o}hlich, J{\"u}rg and Spencer, Thomas},
  journal = {Communications in Mathematical Physics},
  volume  = {84},
  number  = {1},
  pages   = {87--101},
  year    = {1982},
  doi     = {10.1007/BF01208373}
}

@article{Siva2017,
  author = {Siva, Karthik and Yoshida, Beni},
  title = {{Topological order and memory time in marginally-self-correcting quantum memory}},
  journal = {Phys. Rev. A},
  volume = {95},
  number = {3},
  pages = {032324},
  year = {2017},
  month = mar,
  publisher = {American Physical Society},
  doi = {10.1103/PhysRevA.95.032324}
}

@article{Landon-Cardinal2015,
  author = {Landon-Cardinal, Olivier and Yoshida, Beni and Poulin, David and Preskill, John},
  title = {{Perturbative instability of quantum memory based on effective long-range interactions}},
  journal = {Phys. Rev. A},
  volume = {91},
  number = {3},
  pages = {032303},
  year = {2015},
  month = mar,
  publisher = {American Physical Society},
  doi = {10.1103/PhysRevA.91.032303}
}

@article{Rakovszky2023,
  author = {Rakovszky, Tibor and Khemani, Vedika},
  title = {{The Physics of (good) LDPC Codes I. Gauging and dualities}},
  journal = {arXiv},
  year = {2023},
  month = oct,
  eprint = {2310.16032},
}

@article{Nienhuis1982,
  author = {Nienhuis, Bernard},
  title = {{Exact Critical Point and Critical Exponents of $\mathrm{O}(n)$ Models in Two Dimensions}},
  journal = {Phys. Rev. Lett.},
  volume = {49},
  number = {15},
  pages = {1062--1065},
  year = {1982},
  month = oct,
  publisher = {American Physical Society},
  doi = {10.1103/PhysRevLett.49.1062}
}

@article{Madras1990,
  author = {Madras, N. and Orlitsky, A. and Shepp, L. A.},
  title = {{Monte Carlo generation of self-avoiding walks with fixed endpoints and fixed length}},
  journal = {J. Stat. Phys.},
  volume = {58},
  number = {1},
  pages = {159--183},
  year = {1990},
  month = jan,
  issn = {1572-9613},
  publisher = {Kluwer Academic Publishers-Plenum Publishers},
  doi = {10.1007/BF01020290}
}

@article{Bravyi2009nogo,
  author = {Bravyi, Sergey and Terhal, Barbara},
  title = {{A no-go theorem for a two-dimensional self-correcting quantum memory based on stabilizer codes}},
  journal = {New J. Phys.},
  volume = {11},
  number = {4},
  pages = {043029},
  year = {2009},
  month = apr,
  issn = {1367-2630},
  publisher = {IOP Publishing},
  doi = {10.1088/1367-2630/11/4/043029}
}

@article{Yoshida2011,
  author = {Yoshida, Beni},
  title = {{Feasibility of self-correcting quantum memory and thermal stability of topological order}},
  journal = {Ann. Phys.},
  volume = {326},
  number = {10},
  pages = {2566--2633},
  year = {2011},
  month = oct,
  issn = {0003-4916},
  publisher = {Academic Press},
  doi = {10.1016/j.aop.2011.06.001}
}

@article{Zhou2025,
  author = {Zhou, Shu-Tong and Cheng, Meng and Rakovszky, Tibor and von Keyserlingk, Curt and Ellison, Tyler D.},
  title = {{Finite-Temperature Quantum Topological Order in Three Dimensions}},
  journal = {Phys. Rev. Lett.},
  volume = {135},
  number = {4},
  pages = {040402},
  year = {2025},
  month = jul,
  publisher = {American Physical Society},
  doi = {10.1103/n9sq-8cxw}
}

@article{chesi2010self,
  title={Self-correcting quantum memory in a thermal environment},
  author={Chesi, Stefano and R{\"o}thlisberger, Beat and Loss, Daniel},
  journal={Physical Review A—Atomic, Molecular, and Optical Physics},
  volume={82},
  number={2},
  pages={022305},
  year={2010},
  publisher={APS},
  doi = {https://doi.org/10.1103/PhysRevA.82.022305}
}

@article{hutter2012self,
  title={Self-correcting quantum memory with a boundary},
  author={Hutter, Adrian and Wootton, James R and R{\"o}thlisberger, Beat and Loss, Daniel},
  journal={Physical Review A—Atomic, Molecular, and Optical Physics},
  volume={86},
  number={5},
  pages={052340},
  year={2012},
  publisher={APS},
  doi={https://doi.org/10.1103/PhysRevA.86.052340}
}

@article{Brown2016Nov,
  author = {Brown, Benjamin J. and Loss, Daniel and Pachos, Jiannis K. and Self, Chris N. and Wootton, James R.},
  title = {{Quantum memories at finite temperature}},
  journal = {Rev. Mod. Phys.},
  volume = {88},
  number = {4},
  pages = {045005},
  year = {2016},
  month = nov,
  publisher = {American Physical Society},
  doi = {10.1103/RevModPhys.88.045005}
}

@article{wootton2013topological,
  title={Topological phases and self-correcting memories in interacting anyon systems},
  author={Wootton, James R},
  journal={Physical Review A—Atomic, Molecular, and Optical Physics},
  volume={88},
  number={6},
  pages={062312},
  year={2013},
  publisher={APS},
  doi={https://doi.org/10.1103/PhysRevA.88.062312}
}

@article{Haah2011Apr,
	author = {Haah, Jeongwan},
	title = {{Local stabilizer codes in three dimensions without string logical operators}},
	journal = {Phys. Rev. A},
	volume = {83},
	number = {4},
	pages = {042330},
	year = {2011},
	month = apr,
	publisher = {American Physical Society},
	doi = {10.1103/PhysRevA.83.042330}
}

@article{Michnicki2014Sep,
	author = {Michnicki, Kamil P.},
	title = {{3D Topological Quantum Memory with a Power-Law Energy Barrier}},
	journal = {Phys. Rev. Lett.},
	volume = {113},
	number = {13},
	pages = {130501},
	year = {2014},
	month = sep,
	publisher = {American Physical Society},
	doi = {10.1103/PhysRevLett.113.130501}
}

@article{Williamson2024Nov,
	author = {Williamson, Dominic J. and Baspin, Nou{\ifmmode\acute{e}\else\'{e}\fi}dyn},
	title = {{Layer codes}},
	journal = {Nat. Commun.},
	volume = {15},
	number = {9528},
	pages = {9528},
	year = {2024},
	month = nov,
	issn = {2041-1723},
	publisher = {Nature Publishing Group},
	doi = {10.1038/s41467-024-53881-3}
}

@article{Williamson2026Aug,
	author = {Williamson, Dominic J.},
	title = {{Partial Self-Correction in Layer Codes}},
	journal = {Phys. Rev. Lett.},
	volume = {137},
	number = {9},
	pages = {090601},
	year = {2026},
	month = aug,
	publisher = {American Physical Society},
	doi = {10.1103/mb89-8436}
}

@article{Baspin2025Sep,
	author = {Baspin, Nou{\ifmmode\acute{e}\else\'{e}\fi}dyn},
	title = {{The Free Energy Barrier: An Eyring-Polanyi bound for stabilizer Hamiltonians, with applications to quantum error correction}},
	journal = {arXiv},
	year = {2025},
	month = sep,
	eprint = {2509.17356},
}

@article{Dennis2002,
  author = {Dennis, Eric and Kitaev, Alexei and Landahl, Andrew and Preskill, John},
  title = {{Topological quantum memory}},
  journal = {J. Math. Phys.},
  volume = {43},
  number = {9},
  pages = {4452--4505},
  year = {2002},
  month = sep,
  issn = {0022-2488},
  publisher = {AIP Publishing},
  doi = {10.1063/1.1499754}
}

@article{Lu2020,
  author = {Lu, Tsung-Cheng and Hsieh, Timothy H. and Grover, Tarun},
  title = {{Detecting Topological Order at Finite Temperature Using Entanglement Negativity}},
  journal = {Phys. Rev. Lett.},
  volume = {125},
  number = {11},
  pages = {116801},
  year = {2020},
  month = sep,
  publisher = {American Physical Society},
  doi = {10.1103/PhysRevLett.125.116801}
}

@article{Weinstein2019,
  author = {Weinstein, Zack and Ortiz, Gerardo and Nussinov, Zohar},
  title = {{Universality Classes of Stabilizer Code Hamiltonians}},
  journal = {Phys. Rev. Lett.},
  volume = {123},
  number = {23},
  pages = {230503},
  year = {2019},
  month = dec,
  publisher = {American Physical Society},
  doi = {10.1103/PhysRevLett.123.230503}
}

@article{Hamma2009,
  author = {Hamma, Alioscia and Castelnovo, Claudio and Chamon, Claudio},
  title = {{Toric-boson model: Toward a topological quantum memory at finite temperature}},
  journal = {Phys. Rev. B},
  volume = {79},
  number = {24},
  pages = {245122},
  year = {2009},
  month = jun,
  publisher = {American Physical Society},
  doi = {10.1103/PhysRevB.79.245122}
}

@article{Fan2024,
  author = {Fan, Ruihua and Bao, Yimu and Altman, Ehud and Vishwanath, Ashvin},
  title = {{Diagnostics of Mixed-State Topological Order and Breakdown of Quantum Memory}},
  journal = {PRX Quantum},
  volume = {5},
  number = {2},
  pages = {020343},
  year = {2024},
  month = may,
  publisher = {American Physical Society},
  doi = {10.1103/PRXQuantum.5.020343}
}

@article{Hastings2011,
  author = {Hastings, Matthew B.},
  title = {{Topological Order at Nonzero Temperature}},
  journal = {Phys. Rev. Lett.},
  volume = {107},
  number = {21},
  pages = {210501},
  year = {2011},
  month = nov,
  publisher = {American Physical Society},
  doi = {10.1103/PhysRevLett.107.210501}
}

@book{Kardar2007,
  author    = {Mehran Kardar},
  title     = {Statistical Physics of Fields},
  publisher = {Cambridge University Press},
  year      = {2007},
  address   = {Cambridge},
  doi       = {10.1017/CBO9780511815881}
}

@article{Poulin2013,
  author = {Landon-Cardinal, Olivier and Poulin, David},
  title = {{Local Topological Order Inhibits Thermal Stability in 2D}},
  journal = {Phys. Rev. Lett.},
  volume = {110},
  number = {9},
  pages = {090502},
  year = {2013},
  month = feb,
  publisher = {American Physical Society},
  doi = {10.1103/PhysRevLett.110.090502}
}

@article{satzinger2021,
  title={Realizing topologically ordered states on a quantum processor},
  author={Satzinger, Kevin J and Liu, Y-J and Smith, Andras and Knapp, Curt and Newman, Michael and Jones, Cody and Chen, Zhaoshi and Quintana, C{\'e}line and Mi, Xiaohua and Dunsworth, A and others},
  journal={Science},
  volume={374},
  number={6572},
  pages={1237--1241},
  year={2021},
  publisher={American Association for the Advancement of Science},
  doi={https://doi.org/10.1126/science.abi8378}
}

@article{Semeghini2021,
  author  = {Giulia Semeghini and Harry Levine and
             Alexander Keesling and Sepehr Ebadi and
             Tout T. Wang and Dolev Bluvstein and
             Ruben Verresen and Hannes Pichler and
             Marcin Kalinowski and Rhine Samajdar and
             Ahmed Omran and Subir Sachdev and
             Ashvin Vishwanath and Markus Greiner and
             Vladan Vuleti{\'c} and Mikhail D. Lukin},
  title   = {Probing topological spin liquids on a programmable quantum simulator},
  journal = {Science},
  volume  = {374},
  number  = {6572},
  pages   = {1242--1247},
  year    = {2021},
  doi     = {10.1126/science.abi8794}
}

@article{kim2024dynamical,
  title={{Dynamical defects in a two-dimensional Wigner crystal: Self-doping and kinetic magnetism}},
  author={Kim, Kyung-Su and Esterlis, Ilya and Murthy, Chaitanya and Kivelson, Steven A},
  journal={Physical Review B},
  volume={109},
  number={23},
  pages={235130},
  year={2024},
  publisher={APS},
  doi={https://doi.org/10.1103/PhysRevB.109.235130}
}

@article{motta2020determining,
  title={Determining eigenstates and thermal states on a quantum computer using quantum imaginary time evolution},
  author={Motta, Mario and Sun, Chong and Tan, Adrian TK and O’Rourke, Matthew J and Ye, Erika and Minnich, Austin J and Brandao, Fernando GSL and Chan, Garnet Kin-Lic},
  journal={Nature Physics},
  volume={16},
  number={2},
  pages={205--210},
  year={2020},
  publisher={Nature Publishing Group UK, London},
  doi={https://doi.org/10.1038/s41567-019-0704-4}
}

@article{mcardle2019variational,
  title={Variational ansatz-based quantum simulation of imaginary time evolution},
  author={McArdle, Sam and Jones, Tyson and Endo, Suguru and Li, Ying and Benjamin, Simon C and Yuan, Xiao},
  journal={npj Quantum Information},
  volume={5},
  number={1},
  pages={75},
  year={2019},
  publisher={Nature Publishing Group UK, London},
  doi={https://doi.org/10.1038/s41534-019-0187-2}
}

@article{levin2005string,
  title={String-net condensation: A physical mechanism for topological phases},
  author={Levin, Michael A and Wen, Xiao-Gang},
  journal={Physical Review B—Condensed Matter and Materials Physics},
  volume={71},
  number={4},
  pages={045110},
  year={2005},
  publisher={APS},
  doi = {10.1103/PhysRevB.71.045110}
}

@article{griffiths1967correlations,
  title={{Correlations in Ising ferromagnets. I}},
  author={Griffiths, Robert B},
  journal={Journal of Mathematical Physics},
  volume={8},
  number={3},
  pages={478--483},
  year={1967},
  publisher={American Institute of Physics}
}

@article{kelly1968general,
  title={{General Griffiths' inequalities on correlations in Ising ferromagnets}},
  author={Kelly, Douglas G and Sherman, Seymour},
  journal={Journal of Mathematical Physics},
  volume={9},
  number={3},
  pages={466--484},
  year={1968},
  publisher={American Institute of Physics}
}

@article{spohn1999decay,
  title={{Decay of the two-point function in one-dimensional $O(N)$ spin models with long-range interactions}},
  author={Spohn, Herbert and Zwerger, Wilhelm},
  journal={Journal of statistical physics},
  volume={94},
  number={5},
  pages={1037--1043},
  year={1999},
  publisher={Springer},
  doi={https://doi.org/10.1023/A:1004595419419}
}

@article{aoun2021sharp,
	author = {Aoun, Yacine},
	title = {{Sharp asymptotics of correlation functions in the subcritical long-range random-cluster and Potts models}},
	journal = {Electron. Comm. Probab.},
	volume = {26},
	number = {none},
	pages = {1--9},
	year = {2021},
	month = jan,
	issn = {1083-589X},
	publisher = {Institute of Mathematical Statistics and Bernoulli Society},
	doi = {10.1214/21-ECP390}
}

\end{document}